\documentstyle[10pt,epsfig,dp_delphititle]{dp_delphi}
\makeindex
\def\DpPaperGroup{PH-EP}
\def\DpPaperRef{2005-052}
\def\DpDate{2 November 2005}
\def\DpAuthors{DELPHI Collaboration}
\def\DpSubmit{(Accepted by Eur. Phys. J. C)}
\def\DpTitle{
Evidence for an Excess of Soft Photons
in Hadronic Decays of $\mathbf Z^0$
}
\def\DpComment{}
\def\DpEMail{}
\newcommand{\be}{\begin{equation}}
\newcommand{\ee}{\end{equation}}
\newcommand{\bd}{\begin{displaymath}}
\newcommand{\ed}{\end{displaymath}}
\newcommand{\bt}{\begin{tabular}}
\newcommand{\et}{\end{tabular}}
\newcommand{\efig}{\end{figure}}
\newcommand{\bc}{\begin{center}}
\newcommand{\ec}{\end{center}}

\begin{document}
%%%%%%%%%%%%%%%%%%%%%%%%%% They are a problem with Coll.Sty ?
\makeatletter
\makeatother
%%%%%%%%%%%%%%%%%%%%%%%%%% ??????????????????????????????????
% Generate the title page
\begin{titlepage}
\pagenumbering{roman}
\CERNpreprint{\DpPaperGroup}{\DpPaperRef} % Reference of the paper
\date{{\small\DpDate}} % Date of the paper
\title{\DpTitle} % Title of the paper
\address{\DpAuthors} % General name of the author(s)
\begin{shortabs} % Start the abstract
\noindent
%   abstract.tex
%
\noindent
%===================> Abstract     =====> To be filled <=====%
Soft photons inside hadronic jets converted in front of the DELPHI
main tracker (TPC) in events of $q\overline{q}$ disintegrations of the $Z^0$
were studied in the kinematic range $0.2 < E_{\gamma} <$ 1 GeV and 
transverse momentum with respect to the closest jet direction
$p_T < 80$ MeV/$c$. A clear excess of photons in the
experimental data as compared to the Monte Carlo predictions is observed.
This excess (uncorrected for the photon detection efficiency) is 
$(1.17 \pm 0.06 \pm 0.27)\times 10^{-3} \gamma/jet$ in the specified
kinematic region, while the expected level of the inner hadronic
bremsstrahlung (which is not included in the Monte Carlo) is 
$(0.340 \pm 0.001 \pm 0.038)\times 10^{-3} \gamma/jet$.
The ratio of the excess to the predicted bremsstrahlung rate is then
($3.4 \pm 0.2 \pm 0.8$), which is similar in strength to the anomalous soft 
photon signal observed in fixed target experiments with hadronic beams.
\end{shortabs}
\vfill
\begin{center}
\DpSubmit \ \\ % Horrible hack to allow to have DpSubmit empty
\DpComment \ \\
\DpEMail \ \\
\end{center}
\vfill
\clearpage
\headsep 10.0pt
\addtolength{\textheight}{10mm}
\addtolength{\footskip}{-5mm}
\begingroup
% Commands to process the author names
%
\newcommand{\DpName}[2]{\hbox{#1$^{\ref{#2}}$},\hfill}
\newcommand{\DpNameTwo}[3]{\hbox{#1$^{\ref{#2},\ref{#3}}$},\hfill}
\newcommand{\DpNameThree}[4]{\hbox{#1$^{\ref{#2},\ref{#3},\ref{#4}}$},\hfill}
\newskip\Bigfill \Bigfill = 0pt plus 1000fill
\newcommand{\DpNameLast}[2]{\hbox{#1$^{\ref{#2}}$}\hspace{\Bigfill}}
%
%\small
\footnotesize
\noindent
\DpName{J.Abdallah}{LPNHE}
\DpName{P.Abreu}{LIP}
\DpName{W.Adam}{VIENNA}
\DpName{P.Adzic}{DEMOKRITOS}
\DpName{T.Albrecht}{KARLSRUHE}
\DpName{T.Alderweireld}{AIM}
\DpName{R.Alemany-Fernandez}{CERN}
\DpName{T.Allmendinger}{KARLSRUHE}
\DpName{P.P.Allport}{LIVERPOOL}
\DpName{U.Amaldi}{MILANO2}
% ...
\normalsize
\endgroup
\newpage
\titlefoot{Department of Physics and Astronomy, Iowa State
     University, Ames IA 50011-3160, USA
    \label{AMES}}
\titlefoot{Physics Department, Universiteit Antwerpen,
     Universiteitsplein 1, B-2610 Antwerpen, Belgium \\
     \indent~~and IIHE, ULB-VUB,
     Pleinlaan 2, B-1050 Brussels, Belgium \\
     \indent~~and Facult\'e des Sciences,
     Univ. de l'Etat Mons, Av. Maistriau 19, B-7000 Mons, Belgium
    \label{AIM}}
\titlefoot{Physics Laboratory, University of Athens, Solonos Str.
     104, GR-10680 Athens, Greece
    \label{ATHENS}}
\titlefoot{Department of Physics, University of Bergen,
     All\'egaten 55, NO-5007 Bergen, Norway
    \label{BERGEN}}
\titlefoot{Dipartimento di Fisica, Universit\`a di Bologna and INFN,
     Via Irnerio 46, IT-40126 Bologna, Italy
    \label{BOLOGNA}}
\titlefoot{Centro Brasileiro de Pesquisas F\'{\i}sicas, rua Xavier Sigaud 150,
     BR-22290 Rio de Janeiro, Brazil \\
     \indent~~and Depto. de F\'{\i}sica, Pont. Univ. Cat\'olica,
     C.P. 38071 BR-22453 Rio de Janeiro, Brazil \\
     \indent~~and Inst. de F\'{\i}sica, Univ. Estadual do Rio de Janeiro,
     rua S\~{a}o Francisco Xavier 524, Rio de Janeiro, Brazil
    \label{BRASIL}}
\titlefoot{Coll\`ege de France, Lab. de Physique Corpusculaire, IN2P3-CNRS,
     FR-75231 Paris Cedex 05, France
    \label{CDF}}
\titlefoot{CERN, CH-1211 Geneva 23, Switzerland
    \label{CERN}}
\titlefoot{Institut de Recherches Subatomiques, IN2P3 - CNRS/ULP - BP20,
     FR-67037 Strasbourg Cedex, France
    \label{CRN}}
\titlefoot{Now at DESY-Zeuthen, Platanenallee 6, D-15735 Zeuthen, Germany
    \label{DESY}}
\titlefoot{Institute of Nuclear Physics, N.C.S.R. Demokritos,
     P.O. Box 60228, GR-15310 Athens, Greece
    \label{DEMOKRITOS}}
\titlefoot{FZU, Inst. of Phys. of the C.A.S. High Energy Physics Division,
     Na Slovance 2, CZ-180 40, Praha 8, Czech Republic
    \label{FZU}}
\titlefoot{Dipartimento di Fisica, Universit\`a di Genova and INFN,
     Via Dodecaneso 33, IT-16146 Genova, Italy
    \label{GENOVA}}
\titlefoot{Institut des Sciences Nucl\'eaires, IN2P3-CNRS, Universit\'e
     de Grenoble 1, FR-38026 Grenoble Cedex, France
    \label{GRENOBLE}}
\titlefoot{Helsinki Institute of Physics and Department of Physical Sciences,
     P.O. Box 64, FIN-00014 University of Helsinki, 
     \indent~~Finland
    \label{HELSINKI}}
\titlefoot{Joint Institute for Nuclear Research, Dubna, Head Post
     Office, P.O. Box 79, RU-101 000 Moscow, Russian Federation
    \label{JINR}}
\titlefoot{Institut f\"ur Experimentelle Kernphysik,
     Universit\"at Karlsruhe, Postfach 6980, DE-76128 Karlsruhe,
     Germany
    \label{KARLSRUHE}}
\titlefoot{Institute of Nuclear Physics PAN,Ul. Radzikowskiego 152,
     PL-31142 Krakow, Poland
    \label{KRAKOW1}}
\titlefoot{Faculty of Physics and Nuclear Techniques, University of Mining
     and Metallurgy, PL-30055 Krakow, Poland
    \label{KRAKOW2}}
\titlefoot{Universit\'e de Paris-Sud, Lab. de l'Acc\'el\'erateur
     Lin\'eaire, IN2P3-CNRS, B\^{a}t. 200, FR-91405 Orsay Cedex, France
    \label{LAL}}
\titlefoot{School of Physics and Chemistry, University of Lancaster,
     Lancaster LA1 4YB, UK
    \label{LANCASTER}}
\titlefoot{LIP, IST, FCUL - Av. Elias Garcia, 14-$1^{o}$,
     PT-1000 Lisboa Codex, Portugal
    \label{LIP}}
\titlefoot{Department of Physics, University of Liverpool, P.O.
     Box 147, Liverpool L69 3BX, UK
    \label{LIVERPOOL}}
\titlefoot{Dept. of Physics and Astronomy, Kelvin Building,
     University of Glasgow, Glasgow G12 8QQ
    \label{GLASGOW}}
\titlefoot{LPNHE, IN2P3-CNRS, Univ.~Paris VI et VII, Tour 33 (RdC),
     4 place Jussieu, FR-75252 Paris Cedex 05, France
    \label{LPNHE}}
\titlefoot{Department of Physics, University of Lund,
     S\"olvegatan 14, SE-223 63 Lund, Sweden
    \label{LUND}}
\titlefoot{Universit\'e Claude Bernard de Lyon, IPNL, IN2P3-CNRS,
     FR-69622 Villeurbanne Cedex, France
    \label{LYON}}
\titlefoot{Dipartimento di Fisica, Universit\`a di Milano and INFN-MILANO,
     Via Celoria 16, IT-20133 Milan, Italy
    \label{MILANO}}
\titlefoot{Dipartimento di Fisica, Univ. di Milano-Bicocca and
     INFN-MILANO, Piazza della Scienza 2, IT-20126 Milan, Italy
    \label{MILANO2}}
\titlefoot{IPNP of MFF, Charles Univ., Areal MFF,
     V Holesovickach 2, CZ-180 00, Praha 8, Czech Republic
    \label{NC}}
\titlefoot{NIKHEF, Postbus 41882, NL-1009 DB
     Amsterdam, The Netherlands
    \label{NIKHEF}}
\titlefoot{National Technical University, Physics Department,
     Zografou Campus, GR-15773 Athens, Greece
    \label{NTU-ATHENS}}
\titlefoot{Physics Department, University of Oslo, Blindern,
     NO-0316 Oslo, Norway
    \label{OSLO}}
\titlefoot{Dpto. Fisica, Univ. Oviedo, Avda. Calvo Sotelo
     s/n, ES-33007 Oviedo, Spain
    \label{OVIEDO}}
\titlefoot{Department of Physics, University of Oxford,
     Keble Road, Oxford OX1 3RH, UK
    \label{OXFORD}}
\titlefoot{Dipartimento di Fisica, Universit\`a di Padova and
     INFN, Via Marzolo 8, IT-35131 Padua, Italy
    \label{PADOVA}}
\titlefoot{Rutherford Appleton Laboratory, Chilton, Didcot
     OX11 OQX, UK
    \label{RAL}}
\titlefoot{Dipartimento di Fisica, Universit\`a di Roma II and
     INFN, Tor Vergata, IT-00173 Rome, Italy
    \label{ROMA2}}
\titlefoot{Dipartimento di Fisica, Universit\`a di Roma III and
     INFN, Via della Vasca Navale 84, IT-00146 Rome, Italy
    \label{ROMA3}}
\titlefoot{DAPNIA/Service de Physique des Particules,
     CEA-Saclay, FR-91191 Gif-sur-Yvette Cedex, France
    \label{SACLAY}}
\titlefoot{Instituto de Fisica de Cantabria (CSIC-UC), Avda.
     los Castros s/n, ES-39006 Santander, Spain
    \label{SANTANDER}}
\titlefoot{Inst. for High Energy Physics, Serpukov
     P.O. Box 35, Protvino, (Moscow Region), Russian Federation
    \label{SERPUKHOV}}
\titlefoot{J. Stefan Institute, Jamova 39, SI-1000 Ljubljana, Slovenia
     and Laboratory for Astroparticle Physics,\\
     \indent~~Nova Gorica Polytechnic, Kostanjeviska 16a, SI-5000 Nova Gorica, Slovenia, \\
     \indent~~and Department of Physics, University of Ljubljana,
     SI-1000 Ljubljana, Slovenia
    \label{SLOVENIJA}}
\titlefoot{Fysikum, Stockholm University,
     Box 6730, SE-113 85 Stockholm, Sweden
    \label{STOCKHOLM}}
\titlefoot{Dipartimento di Fisica Sperimentale, Universit\`a di
     Torino and INFN, Via P. Giuria 1, IT-10125 Turin, Italy
    \label{TORINO}}
%\titlefoot{INFN,Sezione di Torino, and Dipartimento di Fisica Teorica,
%     Universit\`a di Torino, Via P. Giuria 1,\\
%     \indent~~IT-10125 Turin, Italy
\titlefoot{INFN,Sezione di Torino and Dipartimento di Fisica Teorica,
     Universit\`a di Torino, Via Giuria 1,
     IT-10125 Turin, Italy
    \label{TORINOTH}}
\titlefoot{Dipartimento di Fisica, Universit\`a di Trieste and
     INFN, Via A. Valerio 2, IT-34127 Trieste, Italy \\
     \indent~~and Istituto di Fisica, Universit\`a di Udine,
     IT-33100 Udine, Italy
    \label{TU}}
\titlefoot{Univ. Federal do Rio de Janeiro, C.P. 68528
     Cidade Univ., Ilha do Fund\~ao
     BR-21945-970 Rio de Janeiro, Brazil
    \label{UFRJ}}
\titlefoot{Department of Radiation Sciences, University of
     Uppsala, P.O. Box 535, SE-751 21 Uppsala, Sweden
    \label{UPPSALA}}
\titlefoot{IFIC, Valencia-CSIC, and D.F.A.M.N., U. de Valencia,
     Avda. Dr. Moliner 50, ES-46100 Burjassot (Valencia), Spain
    \label{VALENCIA}}
\titlefoot{Institut f\"ur Hochenergiephysik, \"Osterr. Akad.
     d. Wissensch., Nikolsdorfergasse 18, AT-1050 Vienna, Austria
    \label{VIENNA}}
\titlefoot{Inst. Nuclear Studies and University of Warsaw, Ul.
     Hoza 69, PL-00681 Warsaw, Poland
    \label{WARSZAWA}}
\titlefoot{Fachbereich Physik, University of Wuppertal, Postfach
     100 127, DE-42097 Wuppertal, Germany \\
\noindent
{$^\dagger$~deceased}
    \label{WUPPERTAL}}
\addtolength{\textheight}{-10mm}
\addtolength{\footskip}{5mm}
\clearpage
\headsep 30.0pt
\end{titlepage}
%%%%%%%%%%%%%%%%%%%%%%%%%
%
% Change for the document body
%\pagestyle{heading} % for page numbering
\pagenumbering{arabic} % page numbering in number
\setcounter{footnote}{0} %
\large
%\linenumbers
%   document.tex
%
%*****************************************************************************
\section{Introduction}
Electromagnetic radiation in the soft photon region arising from interacting 
hadrons is assumed to be well understood theoretically. In classical papers by 
Landau and Pomeranchuk \cite{lan} and Low \cite{low} it has been shown that 
the main source of soft photons in hadron reactions 
%in the limit of $E_{\gamma} \rightarrow 0$ 
is the internal hadronic bremsstrahlung, i.e. the bremsstrahlung  radiation 
from the initial and final hadronic states.
Later, Gribov \cite{gribov} defined quantitatively 
the meaning of the term {\em soft photon} as applied to high energy hadron 
reactions: the transverse momentum, $p_T$, of such a photon has to be small
as compared to typical values of this variable for secondary particles 
produced in these reactions, which are 300-400 MeV/$c$.

The experimental investigation of soft photon production in hadron interactions
at high energy started with the bubble chamber experiment \cite{gosh} at SLAC 
in which photons from the reaction  $\pi^- p \rightarrow \gamma + X$  
at an incident momentum of 10.5 GeV/$c$ were studied. A signal consistent 
at the 30\% probability level with the expectations for the inner hadronic 
bremsstrahlung was found. The result was considered as a confirmation of 
theoretical predictions for soft photon production.  
 
However in the next experiment, carried out by the WA27 Collaboration 
at CERN using the BEBC bubble chamber, a clear soft photon signal in excess 
of the inner bremsstrahlung prediction was reported for a $K^+p$ exposure 
at 70 GeV/$c$ \cite{wa27}. After subtraction of photons
coming from all known hadronic decays the residual signal was found to be 
similar in shape to bremsstrahlung but larger in size by a factor of about four 
in the $p_T$ region below 60 MeV/$c$. Then results from the CERN experiment 
NA22 with $K^+$ and $\pi^+$ beams on protons at 250 GeV/$c$ also demonstrated 
an excess of soft photons in a similar kinematic region by a factor of 5 to 7 
as compared to bremsstrahlung predictions \cite{na22}. Similar effects were 
found later in the experiment WA83 where a fine-grained forward electromagnetic 
calorimeter was used to detect photons produced in $\pi^- p$ interactions 
at 280 GeV/$c$ \cite{wa83}, and again in the experiment WA91 which also used 
$\pi^- p$ interactions at 280 GeV/$c$, but implemented a different technique 
for the detection of photons by reconstruction of $e^+e^-$ pairs from photon 
conversion in a thin lead sheet placed in front of the OMEGA spectrometer 
tracker \cite{wa91}. 
 
The situation is less clear with a proton projectile. Experiment E855 at BNL
with protons at 18 GeV/$c$ on Be and W targets did not find 
any signal of direct soft photons at central and slightly backward
rapidities, imposing an upper limit for such photons at 2.7 times the hadronic
bremsstrahlung \cite{bnl}. In a similar photon kinematic region, the experiment 
HELIOS at CERN with proton projectiles at 450 GeV/$c$ on a Be target
found a direct soft photon signal compatible with the expected hadronic
bremsstrahlung, and derived an upper limit on the presence of additional 
sources of direct soft photons of about a factor of two relative to the 
bremsstrahlung \cite{helios}. Recently, the signal of direct soft photons
at the level of four times the bremsstrahlung predictions was observed 
at forward rapidities in $pp$ interactions at 450 GeV/$c$ in the CERN 
experiment WA102 \cite{wa102}. Note, in the latter paper a 
summary of experimental results on direct soft photon observations, 
including kinematic ranges of particular experiments, is also 
given\footnote{In addition, an excess of soft photons with 
$E_\gamma < m_\pi c^2$/2
has been also observed in $\bar{p}p$ interactions at 32 GeV/$c$ 
\cite{mirabel}, however it was not compared with bremsstrahlung predictions.}

Generalizing the results of the experiments which observed an excess
of soft photons, it can be concluded that the photon distributions, 
studied in the very forward region, were reported to be roughly similar 
in shape to that expected for the inner hadronic bremsstrahlung calculated 
from QED, but the observed photon rates were several times larger than 
expected. Owing to this enhancement factor the observed excess photons 
were dubbed $anomalous$.  

Meanwhile various theoretical models [13-33] were suggested, aimed at 
explaining the effect of anomalous photons by introducing new phenomena into 
the soft physics of hadronic interactions. Some of them were able to describe
some particular features of the experimental data, by interpreting 
anomalous soft photons as a radiation from a cold quark-gluon plasma 
\cite{vanhov,liva,lich}, a transient new coherent condition of matter 
\cite{barsh,barsh2,barhei1,barhei2}, or as a synchrotron radiation from quarks 
\cite{nacht1,nacht2,nacht3} in the stochastic QCD vacuum \cite{dosh}. 
However, no model was able to describe the experimental data satisfactorily as 
a whole, especially in a kinematic range where the effect was most prominent
(for a review of the theoretical approaches see \cite{pis,lich}).

In this situation extending the class of reactions in which the phenomenon 
of anomalous soft photons is investigated has become of interest.  
This motivated us to study the reaction 
\begin{equation}
e^+e^- \rightarrow Z^0 \rightarrow hadrons
\end{equation}
at LEP1 searching for extra photons in hadronic decays of $Z^0$ bosons. 

Several studies of photon production in hadronic 
$Z^0$ decays have been carried out by all the LEP experiments 
\cite{opalan,opalq,opalex,insidejet,L3an,L3q,delphian,delphiq,alephq,alephex}
including searches for anomalous photon radiation from non-Standard Model
sources \cite{opalan,L3an,delphian}. The latter aimed at finding 
photons emitted by a non-standard source or by quarks before or
at the beginning of the fragmentation process. Therefore a signal of rather 
hard photons well separated from other tracks was searched for. In contrast, 
the current analysis deals with soft photons deep inside jets, with the aim 
to separate a signal of soft photons coming from the inner hadronic 
bremsstrahlung (mainly from the final hadronic states) or from unknown photon 
sources responsible for the anomalous soft photon radiation seen in hadronic 
experiments. The photon softness can be characterized in this case by a value 
of the transverse momentum of a photon with respect to the closest jet.
We shall use the term $p_T$ for this variable throughout this article.
The $p_T$ range chosen to be studied in this work extends from 0 to 80 MeV/$c$,
while searches for anomalous photons carried out so far in LEP experiments
required them to be {\em hard} and {\em isolated} ($E_{\gamma} > 5$ GeV, 
in general, and at angles to the closest jet $> 20^{\circ}$), thus at 
$p_T > 1.7$ GeV/$c$, i.e. well outside our kinematic region.

This paper is organised as follows. Section 2 provides a description of 
the apparatus, software, and the experimental method applied. This section
also includes a description of selection cuts and data samples. Systematic
uncertainties arising from various elements of the analysis method, 
and their estimates are presented in section 3. 
Section 4 deals with the calculation of the inner hadronic
bremsstrahlung and its systematic errors. In section 5 the main results
of the analysis are given, both uncorrected and corrected for the detection
efficiency. The results show an excess of soft photons in the real data as
compared to the Monte Carlo predictions. Section 6 is devoted to the study 
of possible systematic biases capable of imitating this excess.
Finally, section 7 provides a summary and conclusions.

\section{ Experimental technique and data selection}
\subsection{The DELPHI detector}
The DELPHI detector is described in detail elsewhere \cite{delphi1,delphi2}.
The following is a brief description of the subdetector units relevant
for this analysis. In the DELPHI reference frame the $Z$ axis is taken 
along the direction of the $e^-$ beam. The angle $\Theta$ is the polar
angle defined with respect to the $Z$-axis, $\Phi$ is the azimuthal angle
about this axis and $R$ is the distance from this axis. 

The DELPHI barrel tracking system relied on the Vertex Detector (VD), the 
Inner Detector (ID), the Time Projection Chamber (TPC) and the Outer 
Detector (OD). The barrel electromagnetic calorimeter, the High density
Projection Chamber (HPC) lay immediately outside the tracking detectors.
It was used in this analysis for cross-checks only. All these 
detectors were embedded in a superconducting solenoidal coil providing
a uniform magnetic field of 1.23 T, aligned parallel to the beam axis.

The TPC, the principal device used in this analysis, was the main
tracker of the DELPHI detector; it covered the angular range from  
$20^{\circ}$ to $160^{\circ}$ in $\Theta$ and extended 
from 30 cm to 122 cm in R. It provided up to 16 space points for 
pattern recognition and ionization information extracted from 192 wires.  
%At $cos\Theta=0$ there was a cathode plane which caused a reduced tracking 
%efficiency in the polar angle range $cos\Theta < 0.35$. ???
%The TPC had a two track resolution of about 1.5 cm in $R-\Phi$ and in $Z$.

The HPC covered the angles $\Theta$ from $43^{\circ}$ to $137^{\circ}$.
It had eighteen radiation lengths for perpendicular incidence, and
its energy resolution was $\Delta E/E = 0.31/E^{0.44}\oplus 0.027$ where 
$E$ is in units of GeV \cite{pi0}. It had a high granularity and provided a
sampling of shower energies from nine layers in depth. The angular precisions
for high energy photons were $\pm 1.0$ mrad in $\Theta$ and $\pm1.7$ 
mrad in $\Phi$.  

\subsection{Software}
The principal Monte Carlo (MC) data sets used in this analysis were produced 
with the JETSET 7.3 PS generator \cite{jetset} with parameters adjusted 
according to previous QCD studies \cite{tun1,tun2}. For the test of possible 
systematic biases, two other standard generators: ARIADNE 4.6 \cite{ariadne} 
and HERWIG 5.8C \cite{herwig} with parameters adjusted by the DELPHI tuning 
\cite{tun2} were also used. 

No generation of bremsstrahlung photons from final state hadrons was 
implemented in the MC generators. On the other hand, the initial state 
radiation (ISR) and photon radiation from quarks of $Z^0$ disintegrations
were produced using the DYMU3 generator 
\cite{zitoun} and photon implementation in JETSET \cite{torb}. 
However, as will be shown in section 4, the soft photon rates
from these sources are very small as compared to the bremsstrahlung from
final state hadrons and therefore need not be considered further.

The generated events were fed into the DELPHI detector simulation program 
DELSIM \cite{delphi2} in order to produce data which are as close as possible 
to the real raw data. These data were then treated by the reconstruction and
analysis programs in exactly the same way as the real data.

To reconstruct jets, the LUCLUS code \cite{luclus} with a fixed resolution 
parameter $d_{join}=$ 3 GeV/$c$ was used. To check the stability
of the obtained results, the jet-finding algorithms DURHAM \cite{durham}
and JADE \cite{jade} were also used, both with the resolution parameter
$y_{cut} = 0.01$. The minimal number of jets in the event was required to be 
two.

\subsection{Identification of soft photons}
As has already been said, the anomalous soft photon production 
was  observed in hadronic reactions at small $p_T$ 
and small polar angles relative to an incident hadronic beam. 
In $q\overline{q}$ disintegrations of the $Z^0$ the corresponding 
``beam" direction is represented by the direction of the initial $q$ and/or 
$\overline{q}$ and thus the photon angle $\theta_{\gamma}$ defined with
respect to the parent jet axis is taken as the angular variable in our study, 
with $p_T$ being the photon momentum projected onto the plane perpendicular 
to that axis. The requirement for the kinematic range to correspond to that 
of hadronic reactions (small $p_T$, small $\theta_{\gamma}$ angles) prevents 
the use of the DELPHI electromagnetic calorimeters for the detection 
of soft photons due to the strong pile-up of hard photons hitting 
the calorimeters near the jet axis. 

Fortunately for the aim of this analysis, the DELPHI setup contains a
significant amount of material in front of the sensitive volume of the DELPHI
main tracker, the TPC. About 7\% of all photons in the barrel region are 
converted in front of the tracker. These photons produce in general two 
reconstructible $e^+ e^-$ tracks in the TPC, giving rise to a clean and 
well defined photon sample which is used in this analysis.

The energy threshold for the reconstruction of these photons is 0.2 GeV. 
This, together with the upper cutoff of 1 GeV usually applied in soft photon 
studies, defined the energy range to be investigated, $0.2 < E_{\gamma} < 1$ 
GeV. Since the study of such photons is not typical in the
LEP experiments, we present the characteristics of their detection in detail, 
starting with a description of the algorithm for the reconstruction of 
converted photons from tracks detected in the TPC.

A search was made along each TPC track for points where the tangent of its
trajectory points directly to the beam spot in the $R\Phi$ projection.
Under the assumption that the opening angle of the electron-positron pair 
is zero, this point represented a possible photon conversion point at 
radius $R$. All tracks which have had a solution $R$ that was more than one
standard deviation away from the main vertex, as defined by the beam spot,
were considered to be conversion candidates. If two oppositely charged
conversion candidates were found with compatible conversion point
parameters they were linked together to form the converted photon. The
following selection criteria were imposed:
\begin{itemize}
 \item the $\Phi$ difference between the two conversion points was at most 
30 mrad; 
 \item the difference between the polar angles $\Theta$ of the two tracks
was at most 15 mrad;
 \item at least one of the tracks should have no associated point in front 
of the reconstructed mean conversion radius.  
\end{itemize}
For the pairs fulfilling these criteria a $\chi^2$ was calculated from
$\Delta \Theta, \Delta \Phi$ and the difference of the reconstructed 
conversion radii $\Delta R$ in order
to find the best combinations in cases where there were ambiguous 
associations. A constrained fit was then applied to the electron-positron
pair candidate which forced a common conversion point with zero opening
angle and collinearity between the momentum sum and the line from the
beam spot to the conversion point. 

\subsection{ Photon detection efficiency. Resolutions.}
The photon detection efficiency, i.e. conversion probability combined
with the reconstruction efficiency, was determined with the MC events and 
tabulated against three variables: $E_{\gamma}$,  $\Theta_{\gamma}$, (the
photon polar angle to the beam) and $\theta_{\gamma}$ (the photon polar 
angle to the parent jet axis). The efficiency varies with the energy 
from zero at 0.2 GeV up to 4 - 6\% at 1 GeV, depending on the two other 
variables. Typical dependences of the efficiency on $E_{\gamma}$ and
$\theta_{\gamma}$ are shown in figs. 1a,b. 
%As to the dependence of the efficiency on the $\Theta_{\gamma}$, it reflects 
%the material distribution dependence on this angle.
 
The accuracy of the converted photon energy measurement was found to be about 
$\pm$1.2\% in the given kinematic range. The angular precision of the photon 
direction reconstruction is presented by figs. 1c,d, in which the 
distributions of the difference between the generated and reconstructed 
photon angles $\Theta_{\gamma}$ and $\Phi_{\gamma}$ are shown. These 
distributions have a Breit-Wigner shape, as expected for 
the superposition of many Gaussian distributions of varying width \cite{eadie}. 
The full widths ($\Gamma$'s) of the $\Delta \Theta_{\gamma}$ and 
$\Delta \Phi_{\gamma}$ distributions are 4 and 5 mrad, respectively. 

The importance of good angular resolution in studying anomalous soft photons
was shown in the hadronic beam experiment studies where most of the 
anomalous soft photons were observed inside a cone of 10$-$20 mrad around 
the beam direction \cite{wa91,wa102}. In those experiments the 
angular accuracy was determined by the precision of the measurement of 
the photon polar angle $\theta_{\gamma}$, which varied between 1 to 6 mrad 
(while the accuracy of the beam direction measurement was about 0.1 mrad). 

In hadronic decays of $Z^0$ bosons the accuracy of the measurement of 
the angle between the initial quark direction and the emitted photon 
is determined mainly by the angular accuracy of the reconstruction of the 
former, represented by the jet axis. Typical values of this accuracy 
in two-jet $e^+ e^-$ annihilation events were reported to be 
between 50 and 60 mrad \cite{jet}, depending slightly on the 
jet-finding algorithm (with best results coming from the LUCLUS code). 
These results were tested with DELPHI MC events and  
a similar accuracy was found for the initial quark direction reconstruction. 
Namely, the mean deviation of the reconstructed jet axis from the 
primary quark direction for jets of momenta $>40$ GeV/$c$ was found to be 
about 40 mrad, as can be deduced from the distribution illustrated by fig. 1e, 
and increases up to 50 mrad for smaller jet momenta. This is 
much worse than the corresponding accuracy of hadronic beam 
experiments. The accuracy does not improve by selecting ``good events" 
with small missing energy and/or small missing longitudinal and transverse 
momenta. A rather moderate improvement of the accuracy (to 25 - 30 mrad) 
can be achieved by selecting two-jet events with the jet acollinearity 
smaller than 20 mrad, at the price of a loss of 80\% of the two-jet event 
statistics. No such selections were implemented in this analysis.

Thus the available accuracy of the determination of the initial quark 
direction in this analysis is expected to spread the angular, $p_T$, and 
(most prominently) $p_T^2$ distributions of the possible anomalous
soft photon signal as compared to the experiments with hadronic beams.

\subsection{ Selection cuts and data samples}
Events involving the hadronic decays of the $Z^0$ from the DELPHI data 
of the 1992 to 1995 running periods were used in this analysis. 

Selection of the hadronic events was based on large 
charged multiplicity ($N_{ch} \geq 5$) and high visible energy 
($E_{vis} \geq 0.2 E_{cm}$).
In addition, the condition 
$30^{\circ} \leq \Theta_{thrust} \leq 150^{\circ}$ was imposed, where 
$ \Theta_{thrust}$ is the angle between the thrust axis and the beam direction. These criteria correspond to an efficiency of $(85.2 \pm 0.2) \%$ with a 
$Z^0 \rightarrow \tau^+ \tau^-$ contamination of $(0.4 \pm 0.1)\%$. 

A total of 3,498,655 events of real data (RD) was selected under these cuts 
and confronted with $8\times 10^6$ MC events selected under the same criteria
and properly distributed over all the running periods.

Jets were reconstructed using the detected charged and neutral particles
of the event, the charged particles being selected under the following 
criteria:
\begin{itemize}
\item  $p > 400$ MeV/$c$; 
\item  $\Delta p/p < 100$\%; 
\item  20$^{\circ} \leq \Theta \leq 160^{\circ}$;
\item track length $> 30$ cm;
\item impact parameters below 4 and 10 cm in the $R \Phi$ and $Z$ projections, 
respectively.
\end{itemize}
The neutral particles were taken within the geometrical acceptances of
the subdetectors in which they were reconstructed, within the selection 
criteria of the appropriate subdetector pattern recognition codes 
\cite{delphi1,delphi2}, without additional cuts.

The selection of jets (whatever jet reconstruction algorithm, LUCLUS, DURHAM, 
or JADE having been used) was made with the following cuts:  
\begin{itemize}
 \item 30$^{\circ} \leq \Theta_{jet} \leq 150^{\circ}$; 
 \item $P_{jet} \geq$ 5 GeV/$c$;
 \item no identified electrons (positrons) were allowed in the jets 
(electron identification with a standard DELPHI tag);
 \item if the jet charged multiplicity $N_{ch} = 1$, the charged particle must 
be identified to be not an electron/positron (which is a stronger cut
than the rejection of a particle identified as an electron/positron). 
\end{itemize}
The selection of converted photons was made with the following cuts:
\begin{itemize}
 \item only converted photons with both $e^+, e^-$ arms reconstructed were  
considered;
 \item 20$^{\circ} \leq \Theta_{\gamma} \leq 160^{\circ}$;
 \item 5 cm $\leq R_{conv} \leq 50$ cm, where $R_{conv}$ is the conversion 
radius;
 \item 200 MeV $\leq E_{\gamma} \leq 1$ GeV. 
\end{itemize}
A total of 682,364 converted photons was selected under these cuts in the RD 
and 1,521,030 converted photons in the MC.

\section{Systematic biases and their uncertainties}
In view of the worsening of the signal detectability mentioned in 
section 2.4, the quality of the MC data becomes important and a 
highly accurate simulation of the experimental conditions in the MC stream,
i.e. minimization of systematic effects biasing the MC distributions with
respect to the RD ones, is paramount. Systematic effects due to this bias in 
the MC data are classified into two types: ``software" and ``hardware" 
systematics.

The software systematics are related to an improper reproduction of
experimental spectra of photons and charged particles by the MC event 
generator. The former affects directly the MC produced photon 
distributions, while the latter does this indirectly by biasing the 
reconstructed jet direction. Similar bias can be induced by a jet-finding
algorithm. Estimates for systematic uncertainties of this type are given 
in section 3.2.

The hardware systematics are related to biases in the simulation of 
experimental conditions in the MC stream, i.e. those which appear 
when transporting MC photons through the DELPHI setup 
and reconstructing them (after conversion in
the DELPHI setup material) from hits simulated in the TPC. These features 
have been extensively studied throughout all the LEP1 period, with
necessary corrections being introduced to the MC code. Some details of this 
study can be found in papers \cite{pi0,bstar}. However, in this 
analysis an additional procedure was implemented to improve the simulation 
of experimental conditions in the MC data, called ``recalibration". 

\subsection{Reduction of hardware systematic bias}
The idea was to use wide angle photons ($\theta_{\gamma} > 200$ mrad), for
which no signal of anomalous soft photons is expected, to re-normalize the
material distribution along the photon path in the simulation, and to account 
for possible differences in reconstruction of converted photons from the
TPC hits along $e^+ e^-$ tracks in the MC and RD. Two
types of recalibration were applied. In the first 
one the wide angle soft photons from the MC and RD samples were 
collected into two-dimensional distributions, conversion radius $R_{conv}$   
versus the photon polar angle relative to the beam axis, $\Theta_{\gamma}$.
For the second type of recalibration the photons were
binned according to $E_\gamma$. Bin widths of the 
calibration distributions were varied by factors up to 4 in order
to check the stability of the procedure relative to the binning.   
The distributions, normalized to an equal number of jets passing the 
selection criteria, were used to obtain correction
coefficients in appropriate bins of the above variables. The corrections
were then applied to the MC data. 

Both recalibration procedures were tried and have been found to give 
similar results for the photon rates integrated over the variables used, 
agreeing within the systematic errors discussed
in the next paragraph (see also section 6.4). The 
results were stable relative to the change of the jet-finding algorithm 
(LUCLUS, DURHAM or JADE codes). On the other hand, the 
calibration coefficients varied over LEP1 running periods due to changes 
in the DELPHI detector, i.e. changes of material distribution in front of 
the TPC, e.g. due to upgrade of the VD, etc., which required them
to be found and used individually for each of the running periods. 

To illustrate the quality of the data after recalibration, the distribution of 
the wide angle, $0.2 < E_{\gamma} <1$ GeV photons against the photon polar 
angle with respect to the beam direction, $\Theta_{\gamma}$, is displayed in
fig. 2. The part of the MC data statistically independent from those involved 
in the calculation of the recalibration coefficients is used for this, being
properly distributed over the data-taking periods considered. The averaged
integral difference between the MC and RD in this plot is below 0.9\% (with 
an excess in MC). Expressed in the rate of photons having $p_T < 80$ MeV/$c$ 
(the $p_T$ range under study) which is $18.4 \times 10^{-3} \gamma$/jet, 
the difference is below $0.16 \times 10^{-3} {\gamma}$/jet. This value is used
as an estimate of the systematic errors of hardware origin in the MC data. 
It is quoted in table 1 together with other systematic error estimates 
considered below.

\subsection{Estimation of software systematic errors}
The largest contribution to the software systematic error was found to come 
from the uncertainty for deviations of the MC spectra produced with the event 
generator JETSET 7.3 PS, implemented to obtain the principal MC data sets 
used in this analysis, from the correct distributions of photons in the 
selected kinematic range, $0.2 < E_{\gamma} < 1$ GeV, $p_T < 80$ MeV/$c$. 
These deviations could happen either due 
to an improper description of the QCD processes in this kinematic range 
by the model used in the generator (string fragmentation model, \cite{jetset}),
or due to an inadequate representation of the full set of unstable hadrons
decaying radiatively at the final stage of the hadronization mechanism.

The systematic errors due to the JETSET generator model and its tuning were
estimated in two steps. First, the MC data were used with three different 
tunings described in \cite{tun1,tun2}. In particular, the invariant mass 
cutoff of parton showers $Q_0$, below which partons are not assumed to 
radiate gluons, and which is important in the soft kinematic region, was 
varied between 1.73 and 2.25 GeV/$c^2$ (other parameters correlated with this 
also being varied in order to keep the overall description of the data 
as good as possible)\footnote{At generator level, the stability of the  
soft photon rate was tested in a wider range of the $Q_0$, 
from 0.3 to 2.25 GeV/$c^2$.}. Comparing the photon spectra in our 
kinematic region for all the three tunings, it was found that the integral 
photon rates vary within $\pm 0.4$\% which can be used as an 
estimate of the systematic error due to generator tuning.
Expressed in photon rates, the difference is at the level of 
$0.08 \times 10^{-3} {\gamma}$/jet. 

Then the MC data produced with other commonly used MC generators, ARIADNE and 
HERWIG were studied. The description of a parton shower by ARIADNE is based
on color dipoles \cite{ariadne,ariadne2}, and that of HERWIG on the coherent
parton branching mechanism \cite{herwig}. Unlike the results for high energy 
isolated photons reported in \cite{opalex,L3an,alephex}, no big difference 
between the JETSET and ARIADNE generators was found in the kinematic range
studied in this work. 
As will be shown below (section 6.2), a systematic uncertainty due to the 
generator model for the rate of soft photons of $p_T < 80$ MeV/$c$ can be 
estimated to be at the level of $0.18 \times 10^{-3} {\gamma}$/jet. 
Combining this value with the uncertainty due to the tuning, the systematic 
error due to the event generator was estimated to be   
$0.20 \times 10^{-3} {\gamma}$/jet. This value is quoted in table 1.  
  
A sensitive cross-check of the generator model systematics, making use 
of charged particle spectra (see section 6.5), has shown that the possible
systematic bias of this type is likely to be much less than the quoted errors. 
Another cross-check, based on the comparison of the $\pi^0$ production in the 
MC and RD (see section 6.6) involving both the generator and hardware 
systematics, also demonstrated results in a good agreement with the estimations
above. The cross-checks indicate some overestimation of the systematic errors 
due to the generator model quoted above, nevertheless they are retained.

The next systematic effect to be considered may originate from the   
possible inadequate representation of unstable hadrons decaying radiatively 
(other than $\pi^0$'s) in the MC code, biasing the MC hadron outcome
as compared to the RD. Its study is described in detail in section 6.7. 
It follows from this study that the value of a systematic error due to effects 
of this type is at the level of $0.05 \times 10^{-3} {\gamma}$/jet. 
 
In order to determine the scale of a possible variation of results due to an 
implementation of a jet-finding algorithm other than LUCLUS, the DURHAM and 
JADE algorithms were also used in the analysis (see section 5.1). From the 
variations obtained the systematic error due to the jet finder was derived 
to be at the level of $0.07 \times 10^{-3} \gamma$/jet. 

Since the results of this work are presented both uncorrected and corrected
for the detection efficiency, in the latter case the systematic errors
resulting from the correction have to be taken into account. The integral 
systematic error due to the efficiency correction in the photon $p_T$ 
range below 80 MeV/$c$ was found to be 6\% of the corrected photon rates
(both, for the RD and for the MC, as well as for their difference). This 
error has two components. The first one is an inaccuracy of the efficiency 
determination within the method implemented for this procedure (see section 
2.4), and is equal to 4\%. It was calculated from the MC data by comparing the 
photon $p_T$ distribution taken at the output of the event generator to the 
analogous distribution of reconstructed photons corrected for efficiency, 
the former being taken with the same cuts as the latter. The second component 
of the error above is related to the choice of the variables used to construct 
the efficiency tables. For example, the photon opening angle to the closest 
track can be used instead of the photon polar angle to the parent jet, 
$\theta_{\gamma}$. Another choice of an efficiency table variable could be 
the momentum of the closest track, or the jet charged multiplicity $N_{ch}$ 
(note, all these variables make the efficiency sensitive to the track density 
near the jet axis). These possibilities were tried and indicate the 
uncertainty of this type in the efficiency finding to be about 5\%.  

In a similar way the appropriate systematic errors due to corrections for 
efficiency in individual bins of the photon $p_T$ distribution were found. 

\section{Calculation of the inner bremsstrahlung}
The principal sources of direct soft photons from the reaction (1) are 
expected to be bremsstrahlung from colliding $e^+ e^-$ (initial state 
radiation) and inner bremsstrahlung from final hadronic states. 
%In the soft energy limit \footnote{In this limit one can neglect, 
%in particular, the influence of the ISR on $Z^0$ propagator.} 
For soft photons both source rates can be calculated at once using either of 
two universal formulae: 
\begin{itemize}
\item [i)] the formula, derivable from the Low paper \cite{low} (see  
also \cite{pis,burn}), explicitly displayed for the first time 
in \cite{gosh} and then used by others \cite{wa27,na22,wa83,wa91,wa102}:
\begin{equation}
\frac{dN_{\gamma}}{d^{3}\vec{k}}
=
\frac{\alpha}{(2 \pi)^2} \frac{1}{E_\gamma}
\int d^3 \vec{p}_{1} . . . d^3 \vec{p}_{N}
\sum_{i,j} \eta_{i} \eta_{j}
\frac{- (P_{i} P_{j}) }{ ( P_{i} K )  ( P_{j} K )}
\frac{ d N_{hadrons}}{ d^{3} \vec{p}_{1} ... d^{3} \vec{p}_{N}}
\end{equation}                      

where $K$ and $\vec{k}$ denote photon four- and 
three-momenta, $P$ are four-momenta of beam $e^+, e^-$ and of the $N$ 
charged outgoing hadrons, and $\vec{p}$ are three-momenta of the latter;
$\eta=1$ for the beam electron and for positive outgoing hadrons, 
$\eta=-1$ for the beam positron and negative outgoing hadrons, 
and the sum is extended over all the $N+2$ charged particles involved;
the last factor in the integrand is a differential hadron production rate;
\item [ii)]
the Haissinski formula ~\cite{hais,tim}, which was tested to be 
more stable with respect to lost (undetected) particles and was used 
in \cite{wa83,wa91,wa102}. It has the same form as (2) with 
the scalar products of four-vectors $-(P_i P_j)$ being replaced by
$(\vec{p}_{i \bot} \cdot \vec{p}_{j \bot})$, where
$\vec{p}_{i \bot} = \vec{p}_i-(\vec{n} \cdot \vec{p}_i) \cdot \vec{n}$ and
~$\vec{n}$ is the photon unit vector.
\end{itemize}    

It is known (see \cite{wa91,wa102}), that the two formulae give results in 
complete agreement when used with MC generated particles unaffected by 
detector response, i.e. when all charged particles of an event enter into 
the formulae, with their precise momenta. We have tested the validity of this 
feature for our case in the following way. For every {\em reconstructed} jet 
the parameters of the {\em generated} charged particles lying in the forward 
hemisphere of the jet (including the corresponding beam particle)
were collected and bremsstrahlung distributions for them were calculated, 
with the polar angle of the bremsstrahlung photon to the reconstructed jet 
direction being an angular variable. Note that this method, i.e. usage of 
$generated$ particle momenta while projecting the produced photon onto 
the $reconstructed$ jet direction, is both a) precise and 
b) automatically accounts for the angular resolution of 
the jet direction.  Both formulae gave the same predictions, and these 
results were used in our estimates for the expected bremsstrahlung rates. 
Integrated over our kinematic range, the total bremsstrahlung rate was obtained
to be $17.1 \times 10^{-3} \gamma$/jet; after convolution with the detection
efficiency, it drops to the value of $0.340\times 10^{-3} \gamma$/jet.
Note that the contribution of the ISR to these rates is small, being at the 
level of about 1.5\% of them. The smallness is easy to explain: although the 
ISR from electron/positron beams is much more intense than the ISR from
hadron beams in experiments \cite{wa27,na22,wa83,wa91,wa102}, where it 
contributed a significant amount to the detected photon rate, all the extra 
photons in this experiment are emitted at very small polar angles with respect 
to the beam direction, with the angular distribution peaking 
at $\Theta_{\gamma} = \sqrt{3}/\Gamma$, where $\Gamma$ is a beam 
Lorentz factor ($\Gamma = 0.89 \times 10^5$ at the $Z^0$ peak), thus yielding 
few photons in the barrel region.

The yield of the final state radiation from quarks of $Z^0$ disintegrations
is similarly small. For its estimate the photon implementation in JETSET 
\cite{torb} was used. The $Q_0$ scale introduced for the QED part of the 
shower was varied\footnote{Together with these variations the QCD $Q_0$ scale 
was varied within the range of 0.3 to 2.25 GeV/$c^2$, showing a weak influence 
of this cutoff on the production rate of soft photons off quarks.} down to
its natural lower limit, the constituent quark mass, which is 300 MeV/$c^2$. 
The production rate of photons off quarks in our kinematic range was
found to be at the level of 3\% of the hadronic bremsstrahlung rate\footnote{
The situation changes little when decreasing the QED $Q_0$ cutoff down to
the extreme limit for it, which is about 4 MeV/$c^2$, due to a weak 
(logarithmic) dependence of the quark bremsstrahlung rate on the $Q_0$.}. 
In what follows, neither this nor the ISR yields will be discussed further; 
they are reduced in the RD$-$MC difference and will be ignored.

It follows from \cite{gribov} that the applicability of the formulae above
to the soft bremsstrahlung calculation is restricted in our case ($e^+ e^-$ 
annihilation into hadronic jets at $\sqrt{s} = M_Z$) to the photon kinematic 
domain $p_T << m_{\pi}c$, which is a stronger condition than the 
one mentioned in the introduction for hadronic reactions. 
However, it has been verified (see next paragraph) that the applicability 
holds even at that weaker condition, with an accuracy of about 10\%. 
Nevertheless, the stronger condition is also typically satisfied in our
case since the $p_T$ distribution of calculated inner hadronic bremsstrahlung 
with photons projected onto the plane perpendicular to the initial quark 
direction (i.e. before the spread induced by the angular resolutions) was found 
to peak at 30 MeV/$c$.  

To test the applicability of the formula above the predictions for the 
initial state radiation calculated with this formula were compared with those 
of the DYMU3 generator \cite{zitoun}. For $0.2 < E_{\gamma} < 1$ GeV 
bremsstrahlung photons produced within 100 mrad angles to the $beam$ direction 
($P_T$ to the $beam$ below 100 MeV/$c$) the results coincided within 4\%. For 
the photons produced within 100 mrad angles to a $jet$ direction (the average 
$P_T$ to the $beam$ is 400 MeV/$c$) the difference reached 11\%. Since the 
range of the photon $p_T$ under study in this work is restricted to be within 
80 MeV/$c$, the estimate for the systematic error in the bremsstrahlung 
calculations due to formula (2) appears to be below 4\% of the calculated 
bremsstrahlung rate. This error, together with further contributions to the 
bremsstrahlung calculation uncertainty described below, is given in table 2.

The stability of these calculations was tested using different
event generators (JETSET, ARIADNE and HERWIG) and different jet finders 
(LUCLUS, DURHAM, JADE) obtaining bremsstrahlung rates agreeing within 5\%
when changing the generator (with LUCLUS as jet finder) and 3\% when
changing the jet finder (with JETSET as an event generator), see table 2.   

Finally, when dealing with the results uncorrected for the detection 
efficiency, the generated bremsstrahlung distributions have to be convoluted 
with the efficiency. This induces an additional systematic error of 9\% to 
the bremsstrahlung predictions due to the uncertainty in the efficiency 
determination. This error has two components, similar to those 
discussed at the end of section 3.2. The first one is an inaccuracy of 
the efficiencies within a given determination procedure  
(section 2.4), and is equal to 7\%. It was calculated from the MC data 
by comparing the photon $p_T$ distribution, taken at the output 
of the event generator and convoluted with the detection efficiency, on the 
one hand, to the analogous distribution of reconstructed photons (i.e. at the 
output of the MC stream), on the other hand. The other component of the
error above is related to the choice of efficiency table variables and is
equal to 5\% of the calculated bremsstrahlung rate.

\section{ Experimental results}
\subsection{ Photon distributions. Signal extraction}
The results obtained in this study are presented both uncorrected and corrected
for the photon detection efficiency. However, the principal set of results is
given $uncorrected$ for the efficiency. This is motivated by the fact that
applying efficiency corrections increases both the statistical and systematic
errors of the results. The latter occurs due to an uncertainty in the 
efficiency determination. The former happens because the entries with the 
smallest efficiency, i.e. with the largest weights, dominate the distributions.
% which is equivalent to the statistics loss. 
For example, with the efficiency corrections applied, the softest photons enter
the distributions with weight factors up to one order of magnitude higher than 
those for the photons of moderate energy. Since this article is aimed mainly at
the demonstration of the existence of excess photons, its principal results 
have to be presented with the highest possible statistical and systematic 
accuracy. Therefore, efficiency-corrected results will be given only when the 
absolute photon rates are discussed, namely in section 5.2.  

Thus we start with the $\theta_{\gamma}, p_T$ and $p_T^2$ photon distributions
uncorrected for efficiency. The RD and MC distributions are presented 
in fig. 3 divided by and subtracted from each other. The RD$-$MC distributions
(the right column of panels in fig. 3) are given in units of 
$10^{-3} {\gamma}$/jet, and are accompanied by calculated bremsstrahlung rates.
All the distributions shown demonstrate an excess of soft photons in the RD 
as compared to the MC, and this excess is apparently higher than the expected 
bremsstrahlung level\footnote{There is a systematic excess of the 
bremsstrahlung predictions over the data at the angles $\theta_{\gamma} > 200$ 
mrad. It comes from our recalibration procedure which assumed that no physical
excess of photons exists at these angles. This assumption, invalid in principle
since there exists a certain hadronic bremsstrahlung radiation at wide angles,
induces a small systematic bias to the whole angular range due to an 
overcorrection and consequently lowers the observed photon excess rate. 
However, for the sake of clarity of the presentation we neglect this bias. 
Left neglected, it decreases the signal by an insignificant amount, 
while its accurate treatment would require including the bremsstrahlung
calculation at wide angles into the procedure of the recalibration
which we preferred to avoid here.}. 

To quantify  the excess the difference between the RD and MC was integrated 
in the $p_T$ interval from 0 to 80 MeV/$c$ ($p_T^2 < 0.64 \times 10^{-2}$ 
(GeV/$c$)$^2$), and the value obtained was defined as a signal.
The excess of the RD over the MC as a function of $p_T^2$ was fitted  
by an exponential. The results obtained are: 
\begin{itemize}
\item signal rate 
\begin{equation}
    R_{RD - MC} = (1.17 \pm 0.06 \pm 0.27)\times 10^{-3} \gamma/jet
\end{equation}
while the expected level of the hadronic photon background in this range 
taken from the MC is
\begin{equation}
    R_{MC} = (18.40 \pm 0.04 \pm 0.26)\times 10^{-3} \gamma/jet.
\end{equation}
The calculated level of the inner hadronic bremsstrahlung in the same range 
is, according to section 4,
\begin{equation}
    R_{brems} = (0.340 \pm 0.001 \pm 0.038)\times 10^{-3} \gamma/jet.
\end{equation}
Evaluated in terms of the inner bremsstrahlung rate, the signal is 
$3.4 \pm 0.2 \pm 0.8$. The rates (3) and (5) together with the other ones, 
obtained under various conditions described below, are given in table 3.

 \item the slope of the excess $p_T^2$ distribution (assuming
$dN_{\gamma}/dp_T^2 \sim exp(-B p_T^2)$ for the excess photons) is fitted 
to the value of $B=(251\pm 21)$ (GeV/$c$)$^{-2}$, which is also a good 
estimation for the slope of the inner hadronic bremsstrahlung, but is 
an order of  magnitude steeper than the typical slopes of $p_T^2$ 
distributions of photons in hadronic $Z^0$ decays. 
\end{itemize}

As can be seen from (3) and (5), the relative strength 
of the signal observed (i.e. signal strength expressed in terms of the 
bremsstrahlung rate) is comparable to the amplitudes of the anomalous 
soft photon effects seen in the hadronic beam experiments 
\cite{wa27,na22,wa83,wa91,wa102}. 

In order to check the independence of the signal amplitude on the jet-finding 
algorithm the DURHAM and JADE algorithms were applied to form jets instead of 
LUCLUS. The results were found to agree within the statistical 
errors, i.e. they are stable against the change of the jet-finding algorithm,
see table 3. The LUCLUS to the DURHAM general selection signal ratio 
was found to be $1.10 \pm 0.07$ and the LUCLUS to the JADE ratio was 
$1.09 \pm 0.07$. 

\subsection{ Data corrected for efficiency}
The $\theta_{\gamma}, p_T$ and $p_T^2$ photon distributions
for the data corrected for the efficiency are given in the same form as those
for the uncorrected ones and are displayed in fig. 4. 
%The slope $B$ of the excess $p_T^2$ distribution
%in fig. 4f is $222 \pm 22)$  (GeV/$c$)$^{-2}$. 
The integral signal rate (the RD to MC rate difference integrated over the 
$p_T$ range from 0 to 80 MeV/$c$) is 
$(69.1 \pm 4.5 \pm 15.7)\times 10^{-3} \gamma$/jet and is given in the last
line of table 3. It is about 7\% of the total jet rate, i.e. the absolute 
strength of the signal (the probability to have an excess photon
per jet) is also similar to that found in the hadronic beam 
experiments\footnote{Here a correspondence of the photon production
in a jet (this study) to its production in a minimum bias interaction event
of the hadronic beam experiments is assumed.}. The corresponding inner hadronic 
bremsstrahlung rate is $(17.10 \pm 0.01 \pm 1.21)\times 10^{-3} \gamma$/jet. 

The differential signal rates ($dN_{\gamma}/d p_T$ per 1000 jets) corrected
for efficiency are presented in 10 $p_T$ bins in table 4, together with the 
corresponding predictions for the inner hadronic bremsstrahlung.

\subsection{ Zero signal experiment}
In order to verify the analysis procedure it was applied to the photon 
kinematic domain where the anomalous soft photon excess was highly improbable
(the zero experiment). Such a domain was defined as follows.

Instead of defining the photon kinematic variables with respect to the
parent jet direction, the direction opposite to that of the most distant jet 
was chosen, while the acollinearity between this and the parent jet  
was required to be greater than 200 mrad. Thus, the procedure separates
photons within the jets having an acollinear opposite jet and projects them 
onto the plane perpendicular to the direction of the latter. All other 
elements of the analysis were 
kept untouched, including the calculation of the bremsstrahlung predictions.

The photon distributions obtained with this procedure are shown in fig. 5
for the RD, for the MC, and for their difference. The latter 
distributions agree well with the bremsstrahlung predictions, though due 
to the relatively high statistical errors they are also compatible with 
zero in the range of $p_T < 80$ MeV/$c$. The corresponding photon rates 
are given in table 3.  

The results of the zero experiment have two different applications.
First, they show that no anomalous photons are produced at the very
beginning of the fragmentation process, before the first hard gluon 
emission. Had the photon radiation been produced at this stage (when
two initial quarks are still highly collinear) the signal would be observed 
when relating the photon with the ``antipode" of its parent jet, because 
the antipode jet would memorize the initial direction of the parent quark
(which emits the photon in this scenario), 
unless a hard gluon emission deviates the antipode quark also. 
%Experimental restrictions for such a process will be given in a forthcoming paper.

The other use of the zero experiment is a confirmation of a good
suppression of definite systematic effects, relevant also to our kinematic 
region and capable of producing a spurious excess. These effects 
are mainly of hardware systematics: an underestimation of material amount 
in front of the TPC in the MC code; a global difference in the reconstruction 
of the converted photons in the RD and MC; an improper treatment of background 
hits (noise, cosmics, etc.) in the RD by the pattern recognition program. 
     
A quantitative estimation of possible biases induced by these effects has been
given in section 3.1. Additional tests for these and other systematic 
effects are described in the following section.

\section{Study of systematic biases capable of imitating the observed excess}
\subsection{ Test for external bremsstrahlung}
The most straightforward background capable of imitating the anomalous photon 
signal is the so called external bremsstrahlung, which is the bremsstrahlung 
from electrons (positrons) produced either in (semi)leptonic decays of 
hadrons or by internal or external conversion of 
photons from hadronic decays, when these electrons pass through the 
experimental setup. It also tends to peak at small $p_T^2$, and if it is  
underestimated by the MC, this could lead to an apparent excess of 
soft photons in the RD events. The rejection of jets containing at least one 
electron applied throughout this analysis (see section 2.5) was implemented in 
order to suppress this effect. However, electrons within the jets which escaped 
identification could be, in principle, responsible for the excess observed.

To check this hypothesis, the level of electron admixture in jets was
varied from its natural ratio (dropping the rejection of jets containing 
identified electrons) to a 5 times smaller one, by applying a loose tag for 
the electron identification\footnote {The electron identification in DELPHI 
has different levels of electron tagging. Normally we used the standard tag,
which provides electron identification with efficiency 55\% for electrons
having momenta above 2 GeV/$c$ \cite{delphi2}. The loose tag has a 
higher electron identification efficiency, approaching 80\%.}. 
Thus, if an essential part
of the signal comes from the electron bremsstrahlung, the signal rate should
increase by several times when passing from the maximal rejection
case (with loose electron tag) to the case with no rejection at all.

In fact, no enhancement of the signal rate was found when dropping the
rejection of jets containing identified electrons (see table 3, lines 5,6),  
while the RD and MC rates both changed, by factors of 1.0944 $\pm 0.0055$ 
and 1.0940  $\pm 0.0038$, respectively (the quoted numbers and their 
statistical errors are obtained with photons in 
the $p_T < 80$ MeV/$c$ range). Furthermore, the maximal electron rejection 
(with the loose tag, table 3, lines 7,8) does not decrease the signal, 
which should occur in the case of a contribution to the latter 
due to the external $e^+e^-$ bremsstrahlung (the RD and MC rates decreased 
by factors of 0.9386 $\pm 0.0049$ and 0.9400 $\pm 0.0034$, respectively).

Thus the hypothesis of an extra amount of the external bremsstrahlung from 
electrons inside the jets in the real data as a source of the excess 
appears to be excluded.

\subsection{ Changing MC generator}
In order to check that the observed excess is not an artefact originating
from a particular feature of the implemented  MC generator (JETSET),  
the photon spectra produced with it were compared to those
from ARIADNE. They are plotted in fig. 6. As can be seen from this figure,
there is a rather weak prevalence of JETSET over ARIADNE at $p_T$ below 
80 MeV/$c$. This means that with ARIADNE as an event generator 
the excess of photons would be slightly increased. In amplitude, the observed 
difference is $0.18 \times 10^{-3} \gamma$/jet. Being
expressed in the units of the signal strength, it is less than 15\% in
the photon $p_T$ range below 80 MeV/$c$. This value is used as an 
estimate for the systematic error due to the event generator (section 3.2).

The comparison of JETSET with HERWIG shows a similar feature,
with HERWIG data tending to decrease further the soft photon rate as 
compared to ARIADNE. Thus JETSET appears to be the generator giving the 
maximal soft photon yield among the tested event generators. 

\subsection{ Secondary photons}
When a high energy photon generates an $e^+ e^-$ pair in the material
in front of the TPC the pair particles may radiate bremsstrahlung photons,
which can enter our kinematic region. In most cases such photons have a small
opening angle relative to the parent photon, which leads to a small-angle
enhancement in the distribution of the two-photon opening angles. Such 
enhancements, at angles below 30 mrad, were seen in both the RD and MC 
distributions of the angles between two converted photons, but they 
cancelled in the RD/MC and RD$-$MC distributions. It follows from this that
the given process is well reproduced in the MC stream and cannot be a source  
of the observed excess.

\subsection{ Comment on the pattern recognition bias}  
An important stage of the reconstruction of the converted photons is the
reconstruction of their constituent $e^+$ and $e^-$ tracks from hits left by 
them in the TPC. A possible different treatment of the hits in the MC and RD 
by the pattern recognition would induce a systematic bias to the photon
reconstruction efficiency. This difference may come from numerous sources.
Two of them are listed below: 
\begin{itemize}
\item in the case of the real data the TPC can be loaded by external  
tracks, noise, cosmics, etc., which is difficult to reproduce in the MC stream,
thus resulting in different TPC patterns being fed into the reconstruction 
program of the RD and MC;
\item a difference in the production of true hits from the $e^+$ and $e^-$ 
tracks of the photon conversion in the RD and the MC data may be induced 
by an improper setting of the TPC efficiency in the MC stream 
(e.g. simply due to TPC ageing),   
and may depend on the position of the photon conversion, $e^+e^-$ track lengths 
(which vary with the photon energy) and even on the jet charged multiplicity, 
which produces a varying environment around the $e^+e^-$ track hits.  
\end{itemize}

This difference is expected to be reduced to a great extent by the 
recalibration procedure described in section 3.1. The only possible pattern 
recognition distinction between the RD and MC whose compensation is not ensured 
by this procedure can take place within the range of photon polar angles to 
the parent jet $\theta_{\gamma} \leq 200$ mrad, since the wide angle photons 
($\theta_{\gamma} > 200$ mrad) were used for the recalibration. In this
region the environment around the $e^+e^-$ track hits may be affected by
charged particles of the jet. In such a case the difference in the pattern 
recognition results should depend on the jet charged multiplicity. In order 
to test this possibility the RD to MC ratio was studied in several bands 
of the jet charged multiplicity, 
%(the mean jet charged multiplicity~ $<N_{ch} > = 7$), 
in three angular ranges: $\theta_{\gamma} < 100$ mrad,
$100 \leq \theta_{\gamma} < 200$  mrad, and $200 \leq \theta_{\gamma} < 400$ 
mrad. The ratios obtained for the different $N_{ch}$ bands are in mutual 
agreement (within individual angular ranges) and agree well with the analogous 
global ratios of the RD to MC (table 5). This means that the pattern 
recognition results appear to be the same for the RD and MC within the full 
angular range under consideration.  

\subsection{ Test with charged particles}  
The analysis of photon distributions described in sect. 5.1 was applied to 
artificial photons 
produced from charged pions. The aim of this test was to check that the 
hadronization procedure of the MC event generator in the soft kinematic 
region has no big systematic bias as compared to the analogous process in the 
real data. Being directly implemented for charged particles (which are charged 
pions mainly), it has a straightforward relation to the $\pi^0$ production 
also due to the almost precise SU(2) symmetry of the strong 
interactions\footnote{There are processes which break the SU(2) symmetry
(e.g. decays of $\eta, \eta'$), but their contribution to 
the soft photon rate is small, see sect. 6.7.} 
(earlier the idea of similar tests has been implemented in \cite{gosh,wa27}).
Thus, the method was to take three-momenta of charged particles in the real 
and MC data as a starting point to represent $\pi^0$ distributions, to decay 
these ``$\pi^0$'s" into two photons, to convolute the resulting photon momenta 
with the photon detection efficiency and feed them into the analysis code.  

The resulting distributions are shown in fig. 7. They are similar to 
the genuine photon spectra (cf. fig. 6), however the main result of this test
is an excellent agreement between the RD and MC samples. The
``signal" strength calculated in the same way as that for true photons
is ($0.102 \pm 0.014)\times 10^{-3} \gamma/jet$, i.e. at the level of
9\% of the photon signal (3). Therefore the hadronization mechanism of 
the applied MC code is concluded to work well as far as concerns
the charged pion production. Being related via SU(2)
symmetry with the production of neutral pions, it is expected to reproduce 
it sufficiently well too. The direct comparison of $\pi^0$ production  
in the RD and MC is done in the next section.

\subsection{ $\pi^0$ tests }  
A general and powerful check of the adequacy of the MC data can be done via 
a comparison of the MC and RD $\gamma \gamma$ mass spectra, by 
comparing the $\pi^0$ signals detected in each data set. Note that the 
applicability of the results obtained with the $\pi^0$ tests holds for  
almost the whole soft photon sample under study since the photon production in 
the data is dominated by $\pi^0$ decays, which yield (according to the MC)
almost 92\% of photons in our kinematic range. 

The production of $\pi^0$'s in the DELPHI data of $Z^0$ hadronic decays was 
studied in \cite{pi0} including the $\pi^0$'s arising from two converted
photons. The experimental result for such $\pi^0$'s shows a tendency
for an overestimation of $\pi^0$ production by JETSET 7.3 at low 
$\pi^0$ momenta ($<1$ GeV/$c$). However the results of that work cannot be 
used directly to estimate the systematic errors of the photon background rate 
in our kinematic range. Therefore a special 
analysis of $\pi^0$ production was done in this work to get such an 
estimation from $\pi^0$ signals extracted from the $\gamma \gamma$ mass 
distributions of converted photons. 

The photons (at least two converted photons per jet were required) 
were subdivided into two energy bands: one band of low energy (LE) 
0.2-1 GeV, and one band of higher energy (HE) 1-10 GeV. Each 
HE photon was combined either with a LE photon of a given jet 
or with a HE photon. Both photons in the combination were weighted by the 
recalibration corrections. The $\gamma \gamma$ mass distributions obtained
are shown in fig. 8 for both, the MC and the RD. It can be seen 
from these distributions that there are distortions of the lower-mass parts 
of the $\pi^0$ peaks. They are induced by the external
bremsstrahlung radiation from at least one of the $e^+ e^-$ arms of a
converted photon of the $\pi^0$. Therefore the spectra were fitted
by two Gaussians superimposed over a smooth background, the second Gaussian
being introduced to describe the distortions. The fit results have shown a 
small difference in the MC and RD widths of the first Gaussian, 
$(4.0 \pm 0.1)$ MeV/$c^2$ versus $(4.4\pm 0.1)$ MeV/$c^2$, respectively,
for LE$\times$HE combinations, and $(4.8 \pm 0.1)$ MeV/$c^2$ versus 
$(5.6 \pm 0.1)$ MeV/$c^2$ for HE$\times$HE combinations. 
The $\pi^0$ peak position was stable at 135 MeV/$c^2$, 
as well as the 2nd Gaussian parameters, the widths of the latter being 
13 and 15 MeV/$c^2$ for the LE$\times$HE and HE$\times$HE combinations, 
respectively. The ratio of the integrals under the two Gaussians (with the 
background subtracted) was found to be about two in both cases. The sum 
of the integrals under the Gaussians represents the number of $\pi^0$'s in an 
appropriate $\gamma \gamma$ mass distribution. They are given
\footnote{We do not give the individual Gaussian yields since they interfere 
strongly due to pile-up of the Gaussians, while their sum is close to being 
fit invariant.} in table 6.

From the HE$\times$HE results the HE photon RD to MC ratio was found to be 
$1.020 \pm 0.007$, i.e. the recalibration procedure succeeded in reducing the 
RD and MC difference to the level of 2\% for these photons. For the LE photons
the effect of the recalibration seems to be slightly better, 
the RD to MC ratio deduced from the LE$\times$HE results is $0.986 \pm 0.023$ 
taking into account the factor 1.020 of the HE photon ratio obtained above. 
This agrees well with the recalibration results for the difference residuals
discussed in section 3.1 and illustrated by fig. 2.

Thus the upper limit for the systematic bias of soft photon RD to MC ratio 
which can be obtained from the $\pi^0$ test with converted photons is 1.024 
at the 95\% C.L.
 
In order to get an independent check of this result the whole procedure 
described above was repeated replacing the HE converted photon with a 
calorimetric (HPC) photon in the same energy band and within a 
$\Theta_{\gamma}$ range of $50^{\circ} - 130^{\circ}$. The HPC photons were 
combined with LE and HE converted photons being weighted by the second type 
of the recalibration procedure, i.e. with the energy binning. In spite of a
worse mass resolution (by a factor of 4), the statistical gain due to 
combination of converted photons with those from the HPC was expected to
give a statistical accuracy of the fit results comparable to those
obtained with two converted photon analysis. Since the HPC photons may have
their own systematic bias, the HPC$\times$HPC combinations were also 
involved in the analysis. 

The $\gamma \gamma$ mass distributions for these photon combinations are
displayed in fig. 9. No distortion effects are visible in the distributions 
due to the worse mass resolution and due to a smaller yield of the converted 
photons to the spectra of a) to d) (by a factor of two). The distributions 
were therefore fitted with a single Gaussian superimposed over a smooth 
background. The results of the fit for the numbers of $\pi^0$'s are given 
in table 7. 

It follows from these results that the converted HE photon RD to MC ratio
is $1.014 \pm 0.013$ if one takes into account the proper HPC RD to MC bias
$1.011 \pm 0.004$ obtained from the HPC$\times$HPC signals. This ratio  
agrees well with the double converted photon analysis value $1.020 \pm 0.007$. 
The converted LE photon RD to MC ratio is then $0.985 \pm 0.028$ and 
the upper limit for the systematic bias of the converted soft photon RD to MC 
ratio obtained from this test is 1.031 at the 95\% C.L. 

Thus, the two analyses agree and suggest that there is no excess
of LE photons from $\pi^0$ decays in our kinematic range. A combined upper 
limit for such an excess derived from both analyses is 1.015 at the 95\% C.L. 
This means that the observed soft photon signal is 4 times greater than the 
95\% C.L. upper limit resulting from the identity of the $\pi^0$ production
rates in the RD and MC.

These results, in favour of the absence of any non-negligible systematic bias 
in the current analysis obtained with the $\pi^0$ tests, are of high importance
due to the fact that the photon production in hadronic $Z^0$ events is 
dominated by $\pi^0$ decays, as mentioned above. Small admixtures from 
radiative decays of $B^*$ mesons, $\eta$'s, $\Sigma^0$ baryons and other 
unstable particles to the overall soft photon production rate are not 
significant and have been verified not to change the above conclusion on 
the systematic bias estimations, as discussed in the next section.
   
\subsection{Soft photons from unstable hadrons other than $\pi^0$'s} 
The strongest sources of soft photons from unstable hadrons other than
$\pi^0$'s in hadronic $Z^0$ decays are $B^*$ mesons which have dominant 
radiative branching ratios and low decay momenta. According to the MC, neutral 
and charged $B^*$ mesons yield 3.3\% and 2.7\% of the total soft photon rate
in our kinematic range, respectively.   

In order to check that there is no bias in the DELPHI MC simulation of soft 
photons from the $B^*$ meson decays, the $B$ meson admixture in the data under 
study was varied using the DELPHI B tag \cite{btag}. It was found that the 
signal is stable (within the quoted errors) when varying the $B^*$ photon yield
in our kinematic range by a factor of 40, from 0.3\% with depleted $B^*$ 
production (the anti-B tag applied) to 12\% with enriched $B^*$ production 
(the B tag applied), the corresponding results are presented by lines 9 and 10 
in table 3, respectively.

The reason for this stability is easy to understand. Since the observed soft 
photon signal is of similar strength to the whole relative yield of $B^*$ 
mesons (both are at about 6\% of the total soft photon rate), the DELPHI MC 
would have to be wrong in the prediction of the $B^*$ production rate by about 
100\% to allow the signal to come from these mesons. This is completely 
excluded by the good agreement between the MC and real data for the $B^*$ 
signal amplitude and its characteristics, studied in \cite{bstar}, from which 
the discrepancy between the two data sets is deduced to be below 4\%.    

This fact was used for an independent proof of the statement that an 
improper simulation of soft photons from the $B^*$ meson decays cannot be 
responsible for the signal observed. Making use of the $B^*$ photon yield 
in our kinematic range (6\%, see above) and the experimental agreement of 
the $B^*$ production rate with the simulation, the systematic uncertainty 
in the total soft photon rate due to $B^*$ photons is established to be below 
0.3\% . In absolute value, using the total soft photon production rate (4),
it is less than $0.05 \times 10^{-3} \gamma/$jet. This uncertainty is quoted 
in table 1. 

Similar considerations are applicable to other unstable particles. Therefore  
the yields of $\eta$'s, $\Sigma^0$ baryons and other radiatively decaying
hadrons ($\omega^0, D^*$ mesons, etc.) to our photon kinematic range were 
estimated studying the MC data and published results on their total 
production rates \cite{eta,sig0,dstar}.

It was found from the MC data that the yield of $\eta$ mesons to our 
photon kinematic range is $(1.03 \pm 0.01)$\%, 
the quoted error being statistical. The systematic error 
of this estimate can be obtained by comparing the MC and experimental 
$\eta$ meson total production rates \cite{eta}. They agree within 10\% 
\cite{chliap} which leads to an error in the overall soft photon 
production rate induced by the $\eta$ decays of less than 0.1\%, or below
$0.02 \times 10^{-3} \gamma/$jet. This uncertainty, though relatively small, 
is included into the software systematic error list quoted in table 1.

The photon yields from $\Sigma^0$ baryons, $D^*$ and other unstable hadrons 
to our kinematic range are even smaller due to their lower radiative branching 
ratios and/or higher decay momenta, thus the systematic uncertainties due to 
them can be neglected. 

Finally, the hypothetical situation when the excess photons originate from 
unstable hadrons which are among $Z^0$ decay products, but are not incorporated
(or not incorporated properly) in the implemented MC event generators, was
considered. The method was to calculate the photon $p_T^2$ spectrum from
radiative decays of an {\em a priori} unknown unstable (excited) hadron, the
excitation energy being varied in a wide range (from 35 to 500 
MeV; the mass of the hadron was varied also, from 1 to 5 GeV/$c^2$, 
and was found to affect the results very slightly), and to compare the 
shape of the obtained spectrum with that of the observed excess. In order
to account for a diversity of possible kinematic characteristics of the
assumed hadron (its energy spectrum and angular distribution relative to
a jet) various energy and angular distributions of a large number of unstable 
hadrons were obtained from the DELPHI MC data and used as templates when
generating the results. Given all the needed input parameters, the photon
$p_T^2$ spectra were calculated using two-body decay phase space formulae.

It was found that only very low excitation energies (below 40 MeV) combined
with a narrow angular distribution of the excited hadron are able to produce 
the exponentially decreasing $p_T^2$ spectra similar to that of the 
observed excess (section 5.1). However, no state with such an excitation
energy is present in the PDG tables \cite{pdg}. The nearest candidate for 
such a state is the $B^*$ (with the excitation energy of 46 MeV), but this
state is well incorporated into the DELPHI MC and was directly
tested varying the $B^*$ yield as described above.

From these results the conclusion is drawn that no known hadron decaying 
radiatively can be a source of a viable systematic effect to the observed 
signal.

\section{ Conclusions}
This analysis shows a significant excess of soft photons close to jet axes in 
the hadronic decays of the $Z^0$ collected in the DELPHI experiment at LEP1,
as compared to the Parton Shower MC predictions. The photon kinematic range 
is defined as follows: $0.2 < E_{\gamma} < 1$ GeV, $p_T < 80$ MeV/$c$, the 
$p_T$ being the photon transverse momentum with respect to the parent jet 
direction. The net excess is measured to be 
$(1.17 \pm 0.06 \pm 0.27)\times 10^{-3} \gamma/jet$ for the data uncorrected 
for the photon detection efficiency. This value has to be compared to the
calculated level of the inner hadronic bremsstrahlung which was expected 
to be the dominant source of direct soft photons in this kinematic region 
(but which was not implemented in the standard MC codes used) 
and is obtained to be $(0.340 \pm 0.001 \pm 0.038)\times 10^{-3} \gamma/jet$. 
Expressed in terms of the bremsstrahlung rate, the observed signal is 
$3.4 \pm 0.2 \pm 0.8$.

The various systematic biases capable of producing the excess photons were
carefully studied, leading to the conclusion that the origin of 
the excess cannot be attributed to trivial reasons such as an underestimation 
of the external bremsstrahlung in the MC events, improper simulation of the 
soft photon spectra by an event generator or different treatment of the real 
and MC data by the pattern recognition code. An important point is the good 
agreement between the MC and real data concerning the production and 
detection of $\pi^0$'s when one of the photons of the $\pi^0$ decay is soft.
  
Analogous conclusions can be drawn for the data corrected for the photon
detection efficiency: the observed signal rate is found to be
$(69.1 \pm 4.5 \pm 15.7)\times 10^{-3} \gamma$/jet, while the inner 
bremsstrahlung rate is expected to be 
$(17.10 \pm 0.01 \pm 1.21)\times 10^{-3} \gamma$/jet. Their ratio is 
then $4.0 \pm 0.3 \pm 1.0$. 

The signal amplitudes obtained are close to the anomalous soft photon effects
seen earlier in hadronic reactions at high energy and reported in  
\cite{wa27,na22,wa83,wa91,wa102}. 

\section*{Acknowledgements}
We are grateful to Profs. K. Boreskov, A. Kaidalov, O. Kancheli, L. Okun, 
Yu. Simonov, T. Sj\"ostrand and P. Sonderegger for useful discussions, 
and to Dr. B. French for detailed considerations of several aspects 
of this work.

We are greatly indebted to our technical collaborators, to the members of 
the CERN-SL Division for the excellent performance of the LEP collider, 
and to the funding agencies for their support in building and operating the 
DELPHI detector.\\
We acknowledge in particular the support of\\
Austrian Federal Ministry of Education, Science and Culture, GZ 616.364/2-III/2a,98,\\
FNRS-FWO, Flandres Institute to encourage scientific and technological research in the industry (IWT), Belgium,\\
FINEP, CNPq, CAPES, FUJB and FAPERJ, Brazil,\\
Czech Ministry of Industry and Trade, GA CR 202/99/1362,\\
Commission of the European Communities (DG XII),\\
Direction des Sciences de la Mati\`{e}re, CEA, France,\\
Bundesministerium f\"ur Bildung, Wissenschaft, Forschung und Technologie, Germany,\\
General Secretariat for Research and Technology, Greece,\\
National Science Foundation (NWO) and Foundation for Research on Matter (FOM), The Netherlands,\\
Norwegian Research Council,\\
State Committee for Scientific Research, Poland, SPUB-M/CERN/P03/DZ296/2000,
SPUB-M/CERN/P03/DZ297/2000, 2PO3B 104 19 and 2PO3B 69 23(2002-2004),\\
FCT - Funda\c c\~ao para a Ci\^encia e Tecnologia, Portugal,\\
Vedecka grantova agentura MS SR, Slovakia, Nr. 95/5195/134,\\
Ministry of Science and Technology of the Republic of Slovenia,\\
CICYT, Spain, AEN99-0950 and AEN99-0761,\\
The Swedish Research Council,\\
Particle Physics and Astronomy Research Council, UK,\\
Department of Energy, USA, DE-FC02-01ER41155,\\
EEC RTN contract HPRN-CT-00292-2002.

\newpage
{\bf Table 1.} Systematic uncertainties of the background of hadronic decay 
photons in the range of photon $p_T < 80$ MeV/$c$ for the data uncorrected 
for detection efficiency. The total systematic error is the quadratic sum 
of the individual errors. 

\vskip 0.6cm 
\begin{center}
\begin{tabular}{|c| c| c|}
\hline
&&\\
Source&~Value, $10^{-3}\gamma$/jet~& Percentage of signal rate$^{*)}$\\
&&\\
\hline
\multicolumn{3}{|c|} {Hardware systematics}                          \\
\hline
Material uncertainty       &                   &                     \\
~~and pattern recognition~~&   0.16            & 14                  \\
\hline
\multicolumn{3}{|c|} {Software systematics}                          \\
\hline
Event generator            &   0.20            & 17                  \\
Jet finder                 &   0.07            & 6                   \\
$B^*$ mesons               &   0.05            & 4                   \\
$\eta$ mesons              &   0.02            & 2                   \\
\hline
Total                      &   0.27            & 23                  \\
\hline 
\end{tabular}
\end{center}
$^{*)}$ The signal rate is defined in section 5.1.

\vskip 2cm 
{\bf Table 2.} Systematic uncertainties of the inner hadronic bremsstrahlung
calculations in the range of photon $p_T < 80$ MeV/$c$. The total 
systematic error is the quadratic sum of the individual errors. The
uncertainties, given in absolute photon rates (2nd column of the table) 
correspond to the case of the data uncorrected for detection efficiency.

\vskip 0.6cm 
\begin{center}
\begin{tabular}{|c| c| c|}
\hline
&&\\
 Source    &Value, $10^{-3}\gamma$/jet      & Percentage of brems rate  \\ 
&&\\
\hline 
Formula (2)                        &   0.014   & 4             \\ 
Event generator                    &   0.017   & 5 	       \\ 
Jet finder                         &   0.010   & 3             \\
Convolution with efficiency$^{*)}$ &   0.029   & 9             \\ 
\hline 
Total                              &   0.038   & 11            \\
\hline 
\end{tabular}
\end{center}
$^{*)}$ The systematic error due to convolution with efficiency has to be taken 
into account when dealing with the results uncorrected for efficiency only.  

\newpage
{\bf Table 3.} Signal amplitudes in units of $10^{-3}\gamma$ per jet,
obtained under various selection criteria. The jets satisfy all the selection 
cuts described in section 2.5 and additional cuts (if any), as indicated 
in this table. The jets were formed by the LUCLUS jet-finding code unless 
the DURHAM or JADE codes are referred to explicitly. The errors are 
statistical only. Information on the systematic errors of the experimental 
photon rates and the bremsstrahlung predictions is given in tables 1 and 2, 
respectively. 

\begin{center}
\begin{tabular}{|c| c| c| c|}
\hline
&&&\\
   & Selection conditions           &$~~~~~$Signal$~~~~~$&$~~~~~$Brems$~~~~~$\\
&&&\\
\hline
 1 & General selection                                &1.170$\pm 0.062$& 0.340$\pm 0.001$\\
 2 & General selection, DURHAM                        &1.060$\pm 0.067$& 0.351$\pm 0.001$\\
 3 & General selection, JADE                          &1.070$\pm 0.074$& 0.332$\pm 0.001$\\
 4 &  The zero experiment                             &0.069$\pm 0.048$&0.0750$\pm 0.0002$\\
 5& No rejection of jets containing $e^+,e^-$         &1.170$\pm 0.061$& 0.339$\pm 0.001$\\
 6& No rejection of jets containing $e^+,e^-$, DURHAM &1.050$\pm 0.066$& 0.348$\pm 0.001$\\
 7&Strong rejection of jets with $e^+,e^-$            &1.150$\pm 0.062$& 0.326$\pm 0.001$\\
 8&Strong rejection of jets with $e^+,e^-$, DURHAM    &1.050$\pm 0.067$& 0.336$\pm 0.001$\\
 9& General selection + anti-B tag                    &1.240$\pm 0.167$& 0.363$\pm 0.002$ \\
10& General selection + B tag                         &1.390$\pm 0.159$& 0.326$\pm 0.002$ \\
\hline 
  & General selection, signal corrected for efficiency& 69.1$\pm 4.5~~$& 17.10$\pm 0.01~$\\ 
\hline 
\end{tabular}
\end{center}

\vskip 2cm
{\bf Table 4.} Differential signal and inner hadronic bremsstrahlung rates 
as a function of the photon $p_T$, in units of $10^{-3} \gamma$/jet integrated
over the $p_T$ bin of 8 MeV/$c$ width.
The first errors are statistical, the second ones are systematic.
\begin{center}
\begin{tabular}{| c | c | c |}
\hline
& & \\
$p_T,$ MeV/$c$&$~~$RD$-$MC corrected for efficiency$~~$&$~~~~$Inner hadronic bremsstrahlung$~~~~$\\
& & \\
\hline
  0 - 8  &    $ 0.64\pm 0.38 \pm 0.14  $    & $ 0.685\pm 0.001 \pm 0.048 $   \\
 ~8 - 16 &    $ 2.66\pm 0.84 \pm 0.63  $    & $ 1.584\pm 0.002 \pm 0.112 $   \\
 16 - 24 &    $ 6.48\pm 1.18 \pm 1.46  $    & $ 1.928\pm 0.002 \pm 0.136 $   \\
 24 - 32 &    $ 8.31\pm 1.40 \pm 1.83  $    & $ 2.007\pm 0.002 \pm 0.142 $   \\
 32 - 40 &    $11.01\pm 1.55 \pm 2.46~ $    & $ 1.984\pm 0.002 \pm 0.140 $   \\
 40 - 48 &    $ 8.88\pm 1.69 \pm 2.03  $    & $ 1.926\pm 0.001 \pm 0.136 $   \\
 48 - 56 &    $ 9.70\pm 1.66 \pm 2.25  $    & $ 1.850\pm 0.001 \pm 0.131 $   \\
 56 - 64 &    $ 6.61\pm 1.62 \pm 1.52  $    & $ 1.776\pm 0.001 \pm 0.126 $   \\
 64 - 72 &    $ 7.30\pm 1.60 \pm 1.67  $    & $ 1.704\pm 0.001 \pm 0.121 $   \\
 72 - 80 &    $ 7.58\pm 1.61 \pm 1.76  $    & $ 1.635\pm 0.001 \pm 0.116 $   \\
\hline
\end{tabular}
\end{center}
          
\newpage
{\bf Table 5.} RD to MC ratios in three ranges of $\theta_{\gamma}$ 
as a function of the jet charged multiplicity $N_{ch}$.
\begin{center}
\begin{tabular}{| c | c | c | c |}
\hline
&&&\\
$N_{ch}$ band&$\theta_{\gamma}<$100mrad&100mrad$\leq \theta_{\gamma}<$200mrad&200mrad$\leq \theta_{\gamma}<$400mrad\\
&&&\\
\hline 
1$\leq N_{ch}\leq 3$&$~~~~~1.060\pm 0.013~~~~~$&$1.046\pm 0.010$&0.980$\pm 0.009$\\
3$ < N_{ch}\leq 5$  &$~~~~~1.074\pm 0.010~~~~~$&$1.025\pm 0.007$&1.001$\pm 0.006$\\
5$ < N_{ch}\leq 7$  &$~~~~~1.049\pm 0.009~~~~~$&$1.044\pm 0.007$&0.973$\pm 0.006$\\
$  N_{ch} > $7      &$~~~~~1.067\pm 0.007~~~~~$&$1.028\pm 0.006$&1.005$\pm 0.005$\\
all $ N_{ch}     $  &$~~~~~1.064\pm 0.005~~~~~$&$1.033\pm 0.004$&0.994$\pm 0.003$\\
\hline
\end{tabular}
\end{center}

\vskip 1.8cm
{\bf Table 6.} $\pi^0$ signal amplitudes (numbers of $\pi^0$'s in the $\pi^0$
peaks) obtained with two combinations of converted photons and the upper limits
for the RD/MC ratio of the converted LE photons extracted from the signals.
\begin{center}
\begin{tabular}{| c | c | c |} 
\hline
& & \\
&LE$_{\rm conv}\times$HE$_{\rm conv}$ &HE$_{\rm conv}\times$HE$_{\rm conv}$\\
& & \\
\hline
 RD    &$~~~~~~~~~~~~~~~~~9052 \pm 147~~~~~~~~~~~~~~~~~$&$   19529 \pm 206  $\\	
 MC    &              $   8999 \pm 133$                 &$   18774 \pm 154  $\\	
 RD/MC &              $ ~1.006 \pm 0.022$               &$~~~1.040 \pm 0.013$\\	
\hline 
\multicolumn{3}{|c|}{Resulting RD/MC for the LE photons: 0.986$\pm 0.023$}   \\	
\hline 
\multicolumn{3}{|c|}{                                                   }   \\
\multicolumn{3}{|c|}{$~~~~$Upper limit for the converted LE photon RD/MC ratio (at 95\% C.L.): 1.024 $~~~~$}\\
\multicolumn{3}{|c|}{                                                   }   \\
\hline 
\end{tabular}
\end{center}
\vskip 1.8cm
{\bf Table 7.} $\pi^0$ signal amplitudes obtained with three combinations of
converted and HPC photons and the upper limits for the RD/MC ratio of the 
converted LE photons extracted from the signals. The combined upper limit is 
a summary of both $\pi^0$ analyses (see table 6).
\begin{center}
\begin{tabular}{| c | c | c | c |}
\hline
&&&\\
&LE$_{\rm conv}\times$HE$_{\rm HPC}$&HE$_{\rm conv}\times$HE$_{\rm HPC}$&HE$_{\rm HPC}\times$HE$_{\rm HPC}$\\
&&&\\
\hline 
RD&$~~~~~~~38396 \pm 803~~~~~~~$&$~~~~~~~~91262\pm 905~~~~~~~~$&$~1182370 \pm 5890$\\
 MC    &$   38539 \pm 687  $     &$   89065 \pm 690$  & $ ~ 1156220\pm 4950 $\\
 RD/MC &$~~~0.996 \pm 0.027$     &$~~~1.025 \pm 0.013$& $~~~~~1.023\pm 0.007$\\
\hline 
\multicolumn{4}{|c|}{Resulting RD/MC for the LE photons: 0.985$\pm 0.028$}  \\	
\hline 
\multicolumn{4}{|c|}{                                                      }\\
\multicolumn{4}{|c|}{ Upper limit for the converted LE photon RD/MC ratio (at 95\% C.L.): 1.031}\\
\multicolumn{4}{|c|}{                                                      }\\
\hline 
\multicolumn{4}{|c|}{                                                      }\\
\multicolumn{4}{|c|}{\bf Combined upper limit for the converted LE photon RD/MC ratio: 1.015}\\
\multicolumn{4}{|c|}{                                                      }\\
\hline
\end{tabular}
\end{center}
\newpage
\begin{figure}[1]
\begin{center}
\epsfig{file=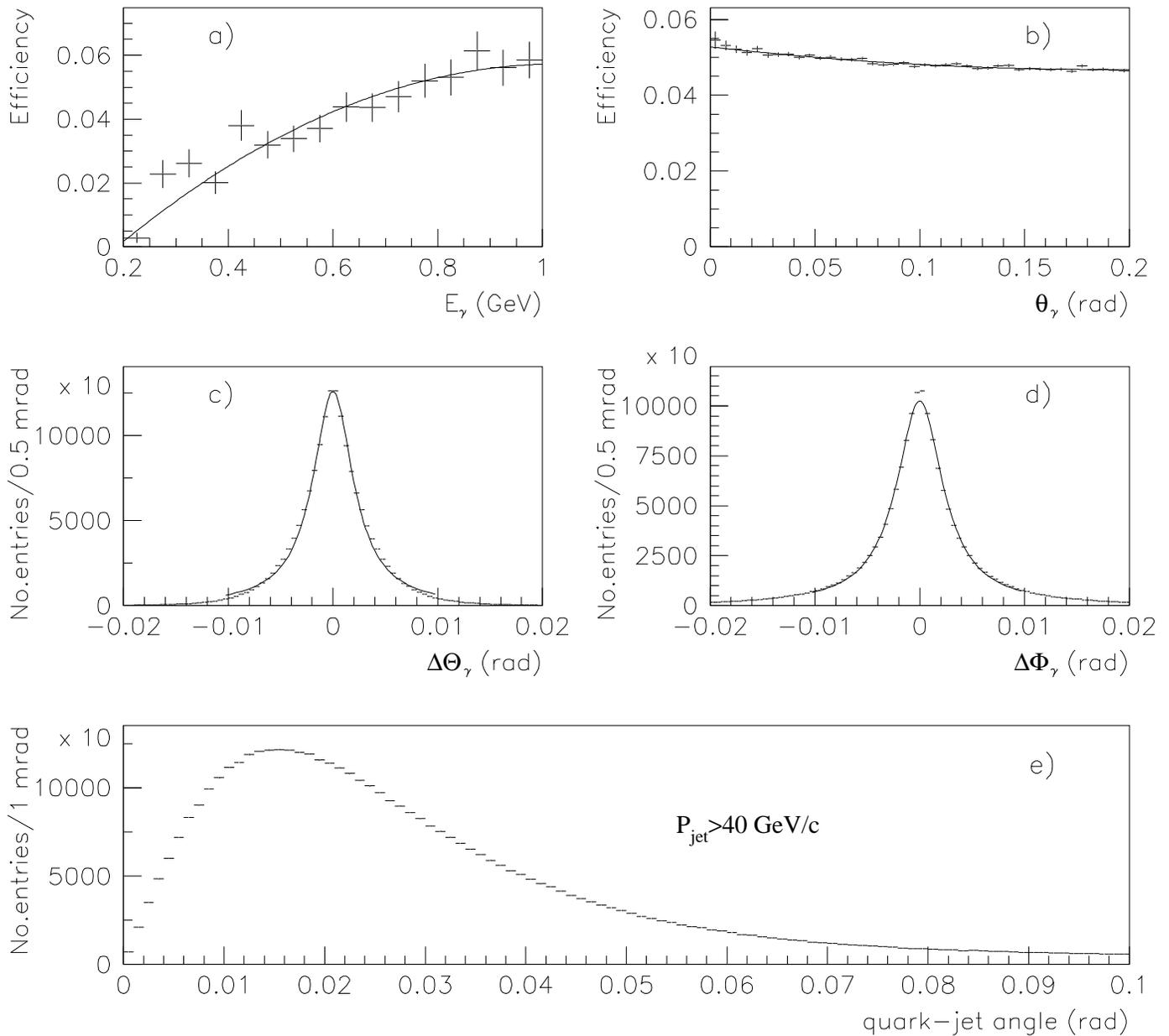,bbllx=50pt,bblly=180pt,bburx=550pt,bbury=570pt,%
width=17cm,angle=0}
\end{center}
\caption{ a) Photon detection efficiency as a function of the photon energy 
in the angular band of $\theta_{\gamma} < 5$ mrad (the statistically poorest 
angular band) integrated over the 3rd efficiency table variable, 
$\Theta_{\gamma}$; b) photon detection
efficiency as a function of $\theta_{\gamma}$ in the $E_{\gamma}$ band 
from 0.9 to 1 GeV (the region of the highest efficiency) integrated over
$\Theta_{\gamma}$; c) difference between generated and reconstructed
photon polar angles $\Theta_{\gamma}$ in the photon energy range of 
$0.2 < E_{\gamma} < 1$ GeV; d) the same for the azimuthal angles 
$\Phi_{\gamma}$; e) deviation of the reconstructed jet axis from the initial 
quark direction for jet momenta $> 40$ GeV/$c$. 
The curves in figs. 1a,b) are 2nd order polynomial fits used for the efficiency 
interpolation. The curves in figs. 1c,d) are the fits by Breit-Wigner form's 
(see text).}
\end{figure}
\newpage
\begin{figure}[2]
\begin{center}
\epsfig{file=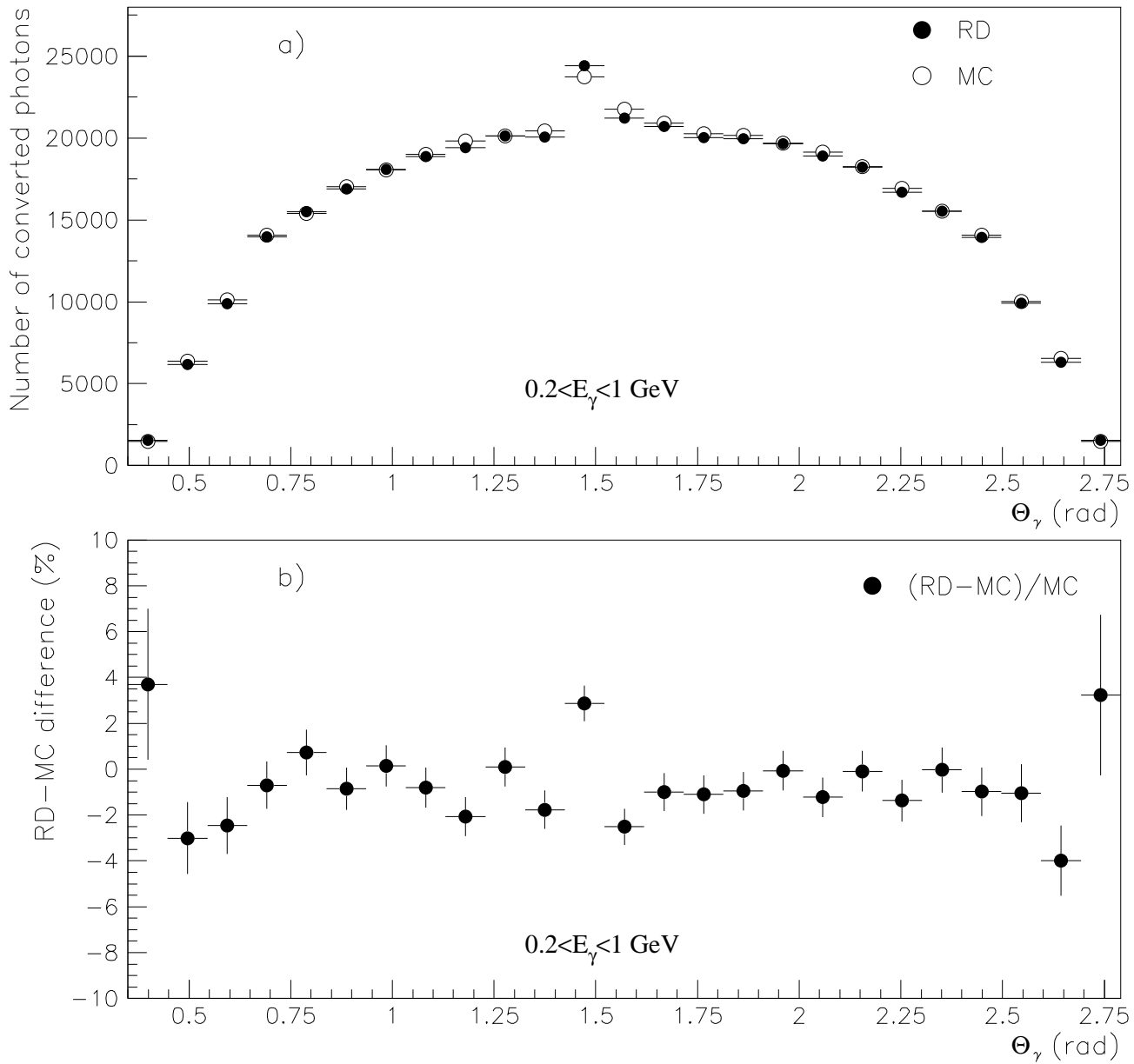,bbllx=50pt,bblly=180pt,bburx=550pt,bbury=570pt,%
width=17cm,angle=0}
\end{center}
\caption{ a) The RD and MC angular distributions (the polar angles 
relative to the beam direction, $\Theta_{\gamma}$) for photons produced 
in hadronic decays of the $Z^0$ and converted in the DELPHI detector before 
the TPC. The photon kinematic range is $0.2 < E_{\gamma} <  1$ GeV and the 
photon polar angle relative to the parent jet direction $\theta_{\gamma} >$ 
200 mrad. The MC data were corrected by the recalibration procedure 
reducing the difference in material distributions in the RD and the MC 
and possible pattern recognition biases (see text); 
b) the relative difference between the RD and corrected MC distributions.}
\end{figure}
\newpage
\begin{figure}[3]
\begin{center}
\epsfig{file=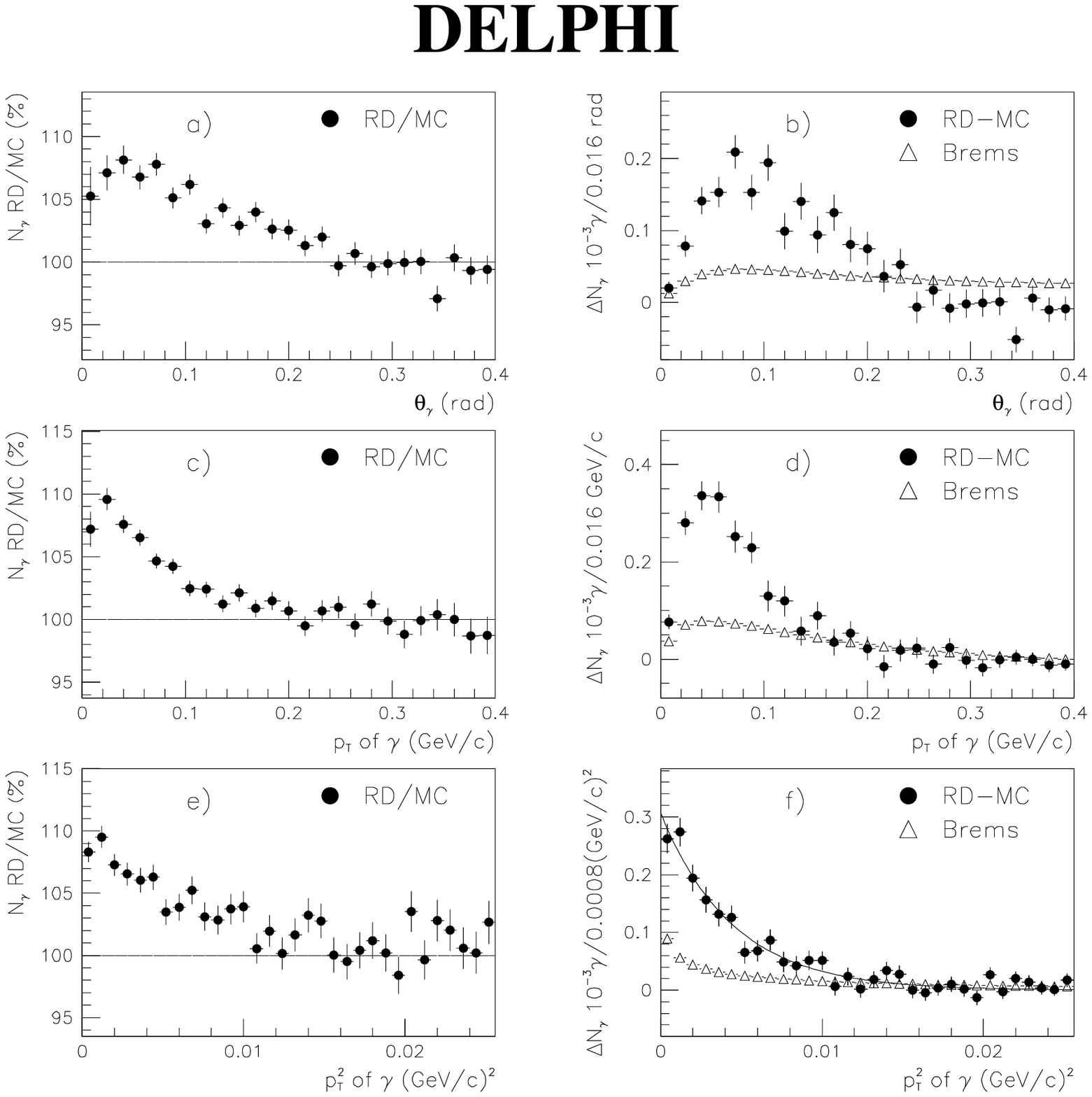,bbllx=50pt,bblly=180pt,bburx=550pt,bbury=550pt,%
width=17cm,angle=0}
\end{center}
\caption{Experimental spectra obtained with the general selection. Left panels: 
the ratio of the RD and MC distributions for a) 
$\theta_{\gamma}$ (photon polar angle relative to the parent jet direction); 
c) photon $p_T$; e) photon $p_T^2$. Right panels, b), d), f): the difference 
between the RD and MC distributions for the same variables, respectively. 
``Brems" corresponds to the inner hadronic bremsstrahlung predictions. 
The errors are statistical. The curve in fig. 3f) is the fit by an exponential
(see text).}
\end{figure}
\begin{figure}[4]
\begin{center}
\epsfig{file=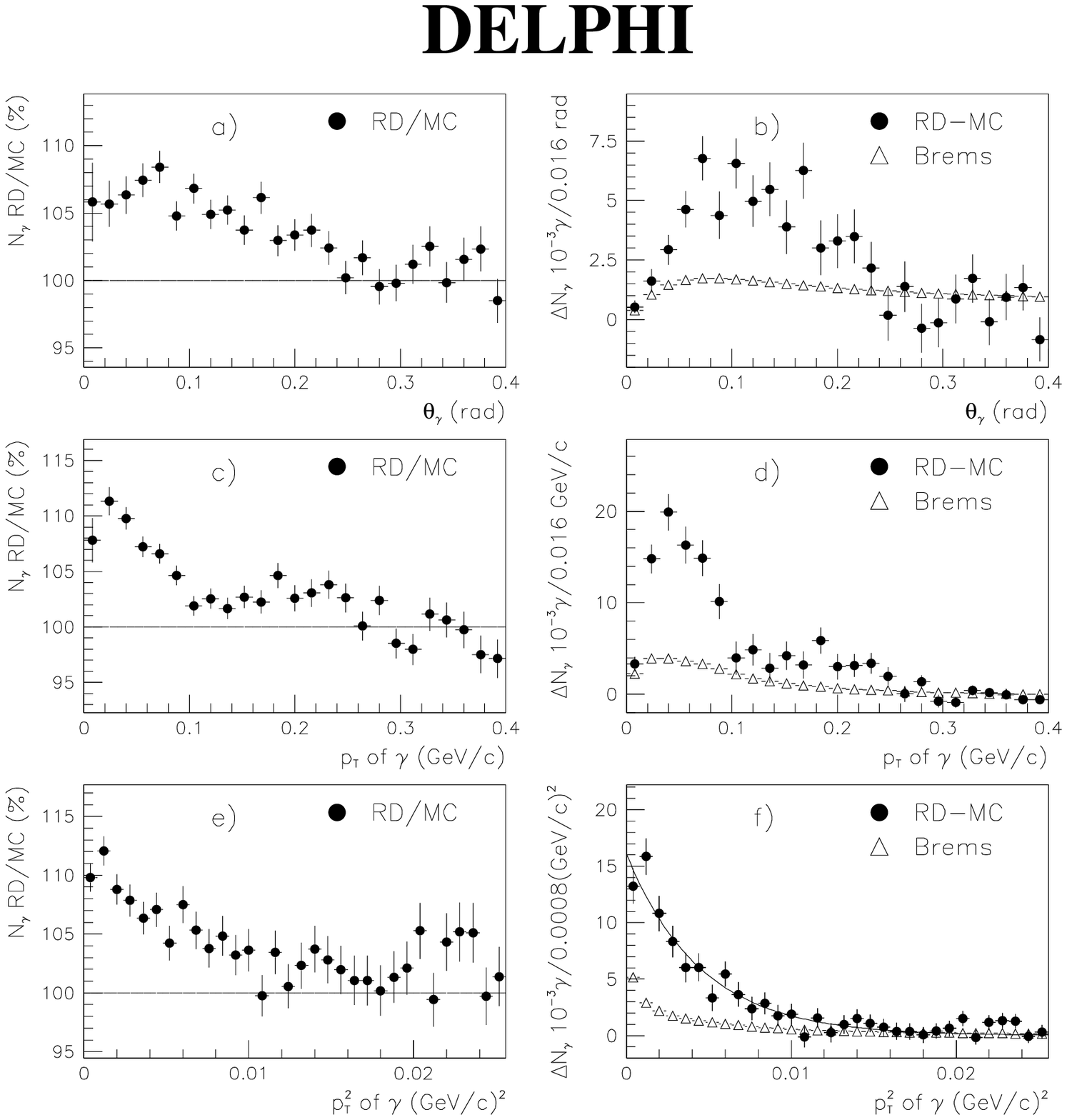,bbllx=50pt,bblly=180pt,bburx=550pt,bbury=550pt,%
width=17cm,angle=0}
\end{center}
\caption{The same as in fig. 3, corrected for the efficiency of photon
detection.}
\end{figure}
\newpage
\begin{figure}[5]
\begin{center}
\epsfig{file=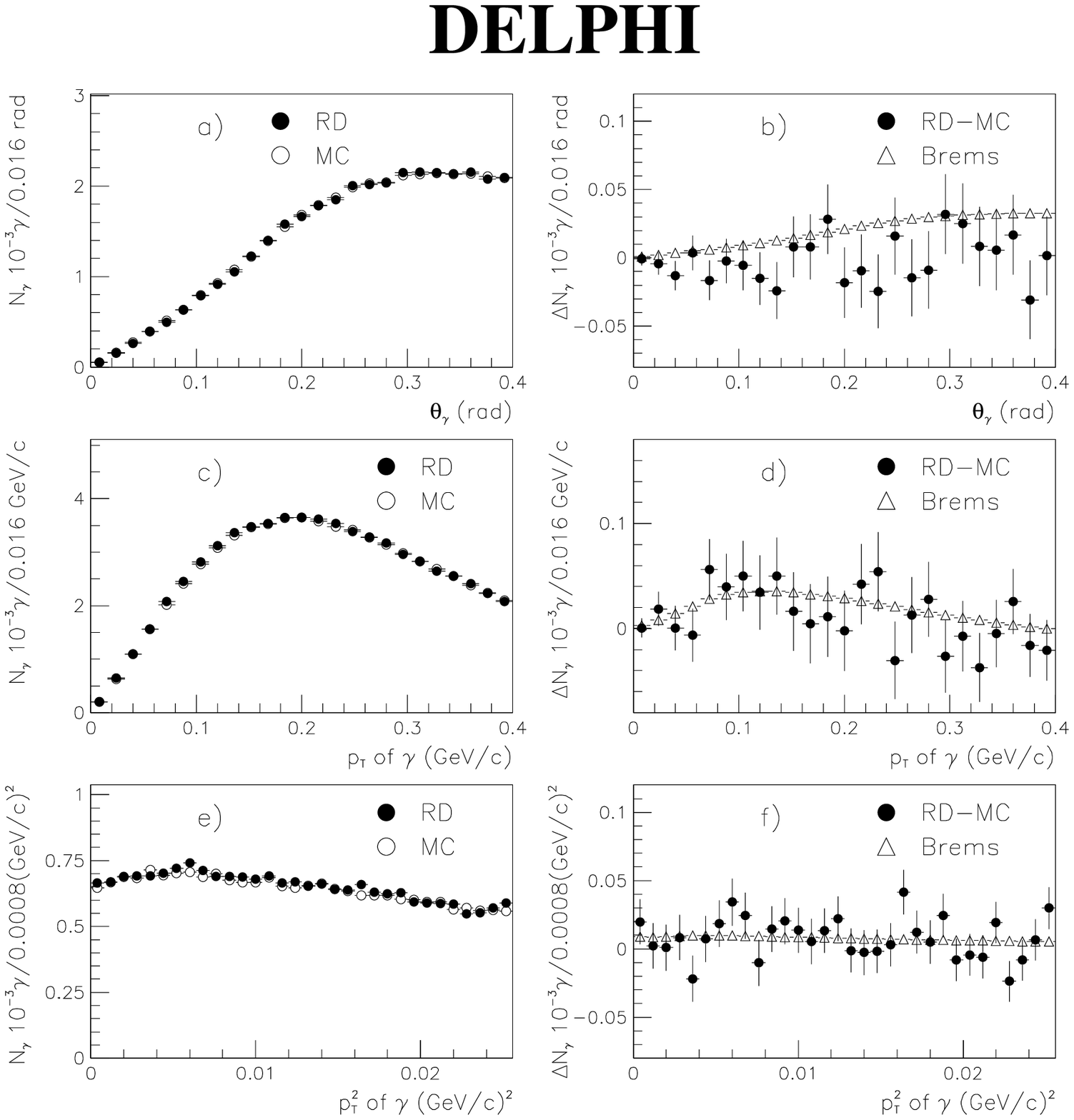,bbllx=50pt,bblly=150pt,bburx=550pt,bbury=500pt,%
width=17cm,angle=0}
\end{center}
\caption{ Zero experiment photon distributions. Left panels, a), c), e): 
the RD and MC $\theta_{\gamma}$, $p_T$ and $p_T^2$ distributions; right
panels, b), d), f): the difference between the RD and MC distributions 
for the same variables together with the bremsstrahlung predictions. 
The errors are statistical.} 
\end{figure}
\newpage
\begin{figure}[6]
\begin{center}
\epsfig{file=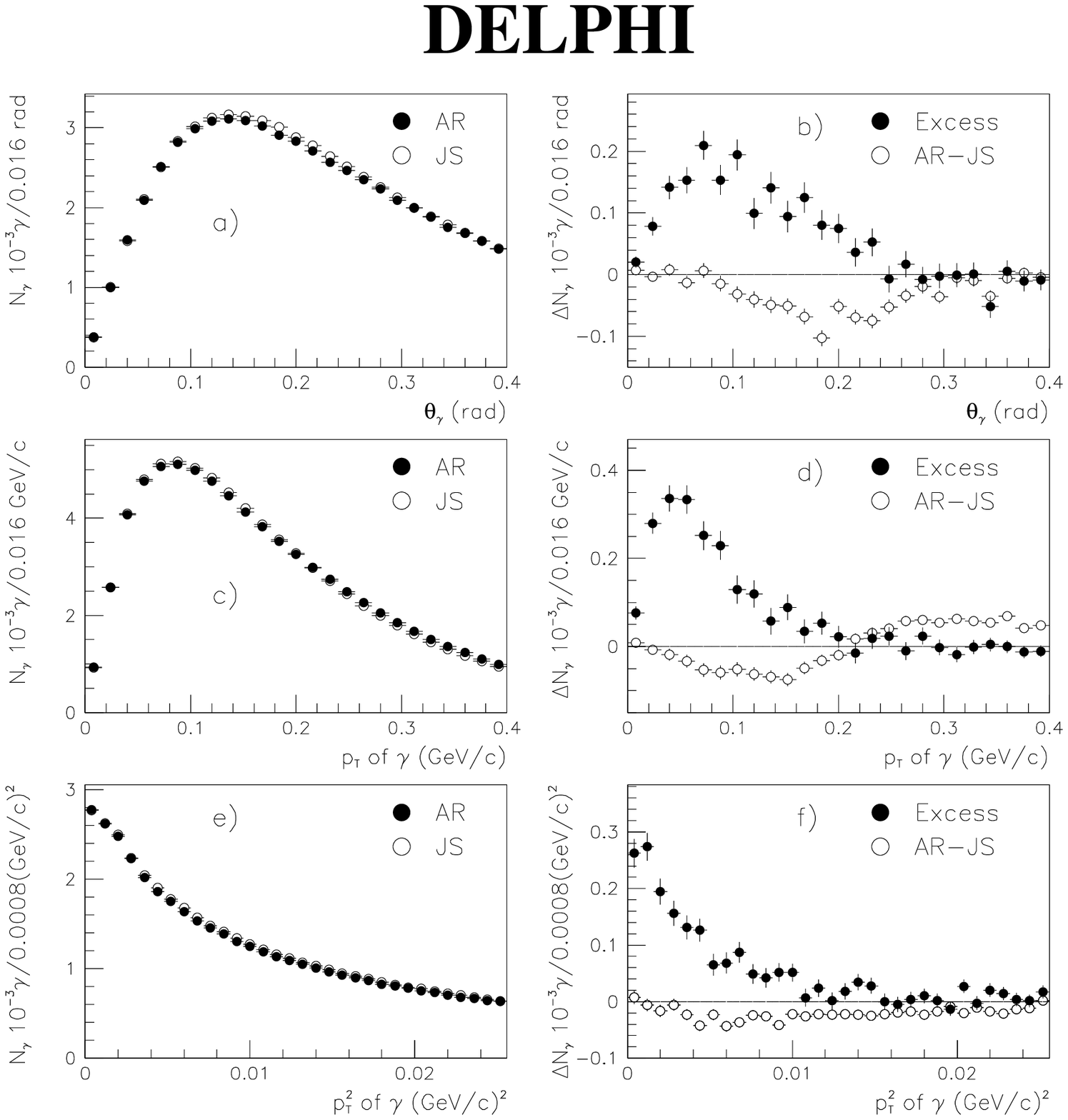,bbllx=50pt,bblly=150pt,bburx=550pt,bbury=550pt,%
width=17cm,angle=0}
\end{center}
\caption{Comparison of the JETSET (JS) and ARIADNE (AR) generators. 
Left panels, a), c), e): the MC $\theta_{\gamma}$, $p_T$ and $p_T^2$
distributions of photons as produced by JETSET and by 
ARIADNE. Right panels, b), d), f): the difference between the two MC 
distributions for the same variables (open circles). The RD$-$MC distributions 
from fig. 3 presenting the observed excess are also displayed for comparison.
The errors are statistical.} 
\end{figure}
\newpage
\begin{figure}[7]
\begin{center}
\epsfig{file=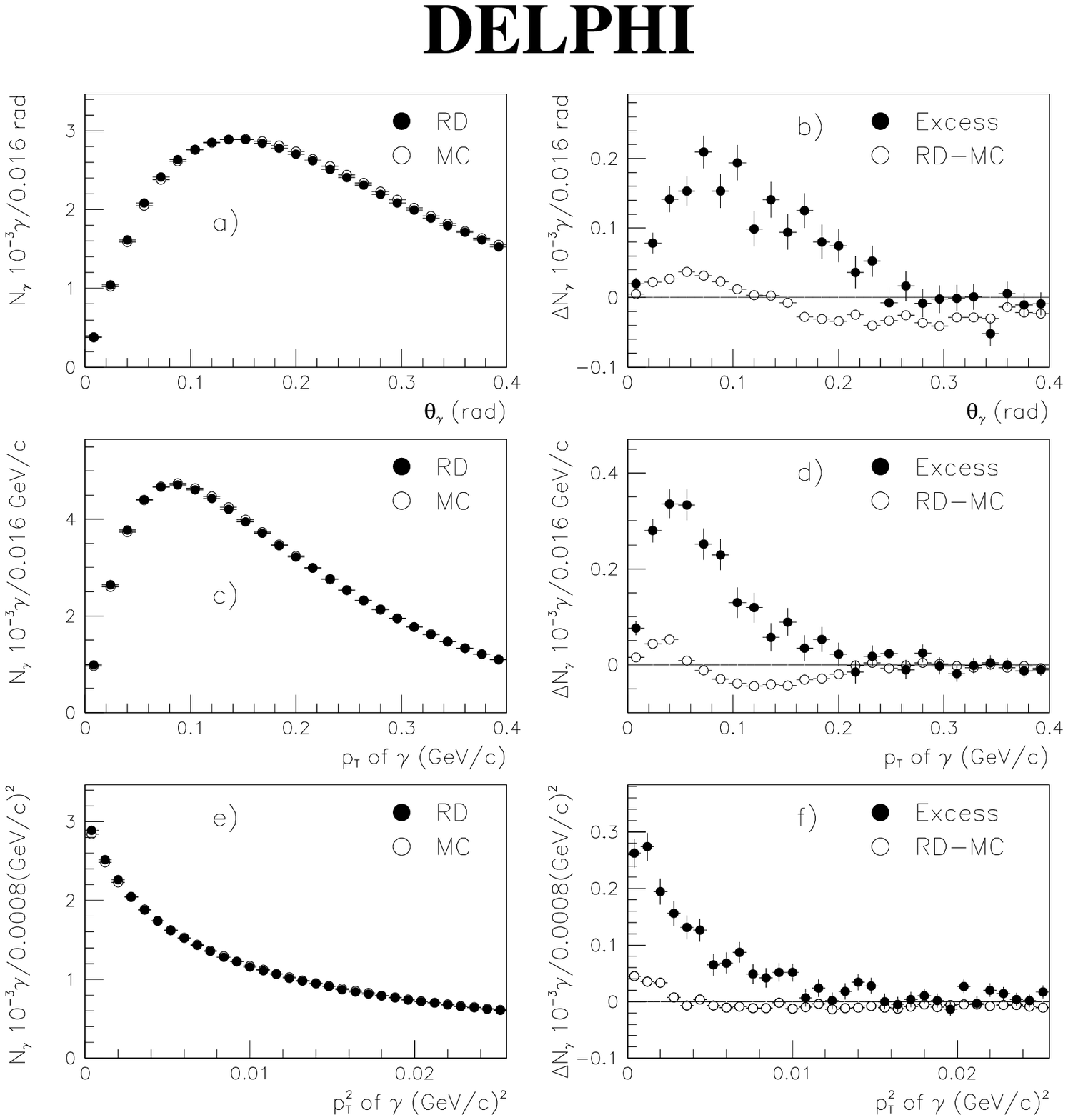,bbllx=50pt,bblly=150pt,bburx=550pt,bbury=550pt,%
width=17cm,angle=0}
\end{center}
\caption{Comparison of the RD and MC distributions of ``photons" produced 
from charged particles (see text, section 6.5). The RD$-$MC distributions for 
these ``photons" are shown by open circles. The observed excess distributions
(RD$-$MC from fig. 3) are also displayed for comparison. 
The errors are statistical.} 
\end{figure}
\newpage
\begin{figure}[8]
\begin{center}
\epsfig{file=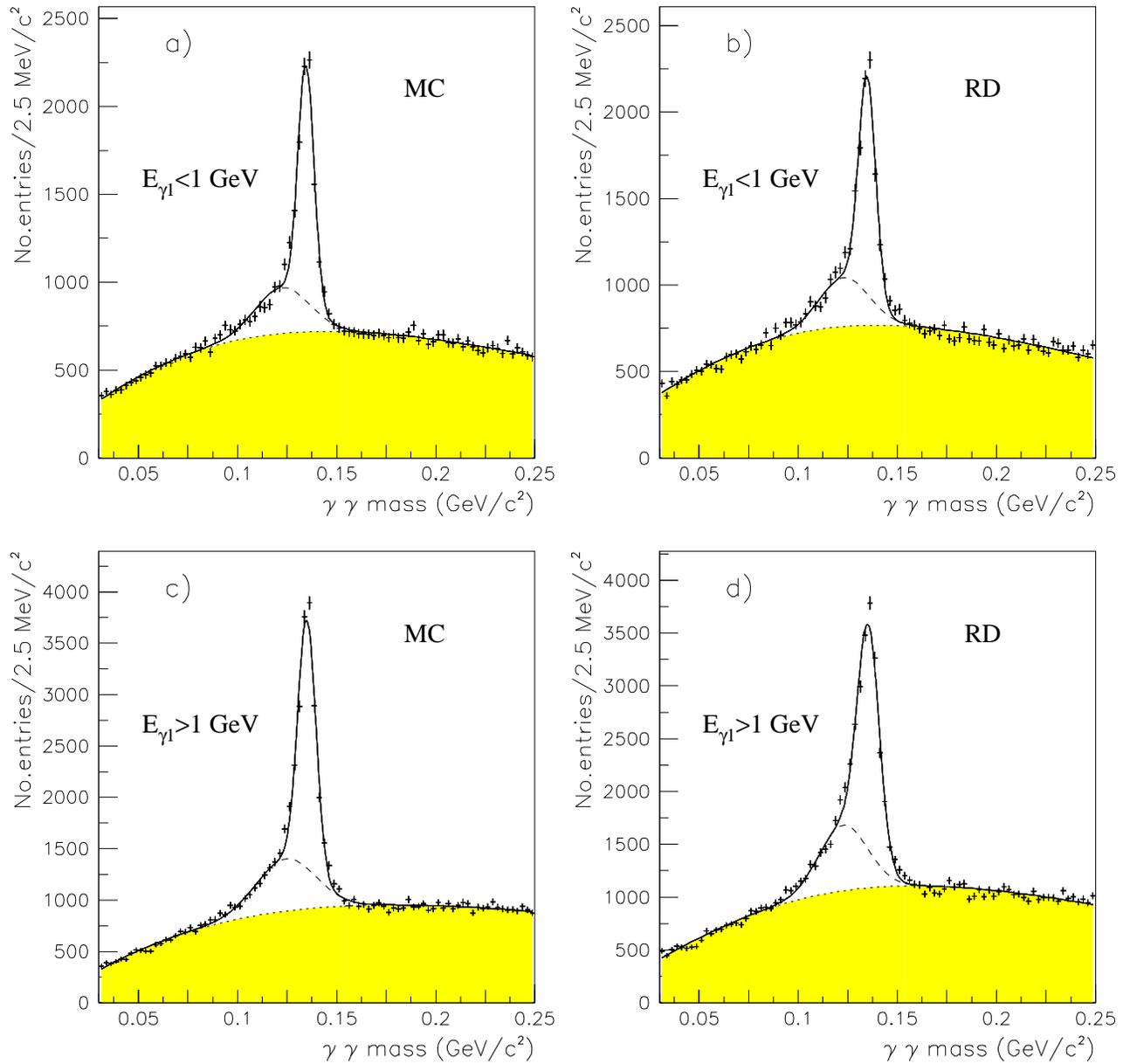,bbllx=50pt,bblly=150pt,bburx=550pt,bbury=550pt,%
width=17cm,angle=0}
\end{center}
\caption{ Comparison of the MC and RD $\gamma \gamma$  mass distributions for 
the two converted photon combinations. a,b) LE$\times$HE combination; c,d) 
HE$\times$HE combination. The dashed line represents the distortion
Gaussian (see text). The errors are statistical. The results of the comparison 
are given in table 6.} 
\end{figure}
\newpage
\begin{figure}[9]
\begin{center}
\epsfig{file=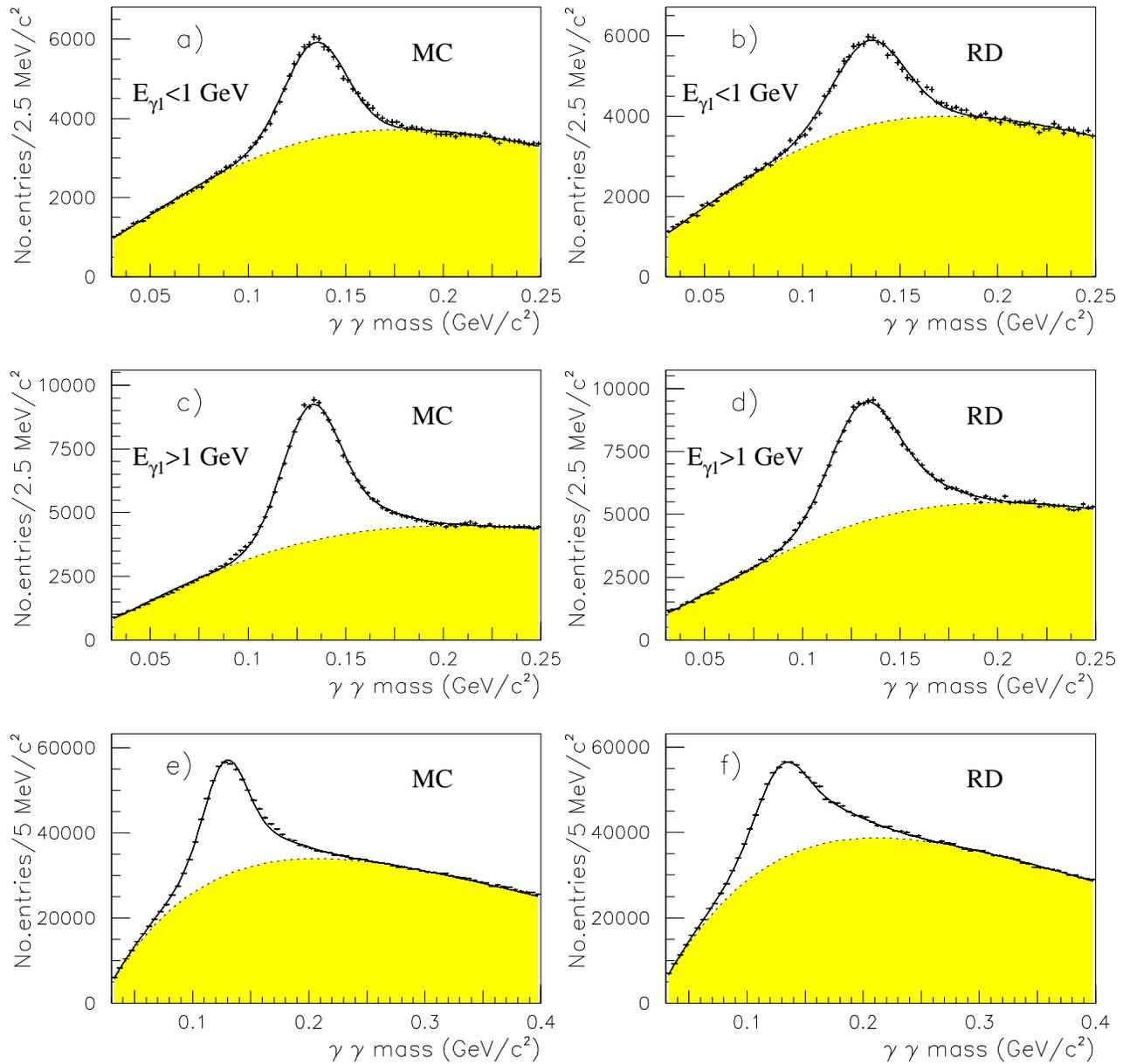,bbllx=50pt,bblly=150pt,bburx=550pt,bbury=550pt,%
width=17cm,angle=0}
\end{center}
\caption{ Comparison of the MC and RD $\gamma \gamma$  mass distributions for 
converted and high energy HPC photon combinations. a,b) converted 
LE and HPC combination; c,d) converted HE and HPC combination;
e,f) HPC and HPC combination (note the different mass scale for these plots).
The errors are statistical. The results of the comparison are given in 
table 7.} 
\end{figure}
\end{document}